\documentclass[a4paper,11pt]{article}
\pdfoutput=1 
\usepackage{jcappub} 
\usepackage{color,soul}
\usepackage{float}
\usepackage{appendix}
\usepackage{subfigure}

\title{ Study of warm inflationary models and their parameter estimation from CMB
}

\author[a,b]{Richa Arya,}
\author[c]{Raghavan Rangarajan}	
\affiliation[a]{Theoretical Physics Division, Physical Research Laboratory, Ahmedabad 380009, India}
\affiliation[b]{Indian Institute of Technology Gandhinagar, Palaj, Gandhinagar 382355, India}
\affiliation[c]{School of Arts and Sciences, Ahmedabad University, Ahmedabad 380009, India}
	
\emailAdd{richaarya@prl.res.in,\, raghavan@ahduni.edu.in}

\abstract{Observations of the temperature anisotropies in the Cosmic Microwave Background (CMB) radiation show that the models of inflation with the monomial potentials are inconsistent with the allowed $n_s-r$ bounds. However certain monomial potentials of inflation are allowed in the context of \textit{Warm Inflation}, where the inflaton's coupling with other fields are significant both \textit{during and after} the inflationary phase. In our study, we consider $\lambda\phi^4$ and $\lambda\phi^6$ models of warm inflation with different forms of the inflaton dissipation coefficient. We parameterize the primordial power spectrum in terms of the model parameters, namely, the inflaton self coupling, $\lambda$, and the dissipation parameter, $Q_P$, due to inflaton's interaction with the other fields. Then we obtain the joint and marginal distributions of these parameters by carrying out a Markov Chain Monte Carlo (MCMC) analysis using the {\tt CosmoMC} numerical code. An estimation of these physical parameters is essential for model building. We also obtain the $n_s$ and $r$ values for the mean values of the parameters and find them to be consistent with the observational bounds, confirming that these simple models are viable models for describing inflation. }

\newcommand{\be}{\begin{equation}}
\newcommand{\ee}{\end{equation}}

\begin{document}
\maketitle
\flushbottom

\section{Introduction}
\label{sec:intro}

Cosmological Inflation \cite{Starobinsky:1980te,Kazanas:1980tx,Sato:1980yn,Guth:1980zm,Linde:1981mu,Albrecht:1982wi,Linde:1983gd} is a phase of accelerated expansion for a very brief duration in the early Universe. It provides a solution to many shortcomings of the Standard Big Bang Model of cosmology. As a bonus, it also provides a mechanism to generate the primordial density fluctuations that can explain the anisotropies in the Cosmic Microwave Background (CMB) radiation and seed the growth of Large Scale Structure (LSS) at late times. (For a review, refer to Refs. \cite{Baumann:2009ds,Riotto:2002yw}.)

\vspace{0.2cm} 
Inflation is driven by a scalar field, $\phi$, known as the \textit{inflaton}. When the energy density of the inflaton field dominates the energy density of the Universe, inflation takes place. In the standard  description, one presumes that the inflaton's coupling to other fields are ineffective during the slow roll inflationary phase. During inflation, the Universe expands nearly exponentially and consequently, the number densities of all species dilute away and the Universe enters into a supercooled state. When inflation ends, the Universe undergoes a \textit{reheating} phase in which the inflaton oscillates and dissipates its energy into particles \cite{Kofman:1994rk}. We refer to this inflationary scenario as {\it Cold Inflation.}

\vspace{0.2cm}
 However, there is another description of inflation known as \textit{Warm Inflation} \cite{Berera:1995wh,Berera:1995ie,Berera:1998px}, in which one considers the inflaton's coupling with other fields  both \textit{during and after} inflation. As the inflaton dissipates into radiation during the slow roll inflationary phase as well, the Universe is not supercooled and has a temperature and is hence \textit{warm} during inflation (for a review refer to Refs. \cite{Berera:2008ar,Rangarajan:2018tte}). The condition for warm inflation to take place is that the temperature of the thermal bath of radiation should be greater than the Hubble expansion of the Universe ($T>H$), so that the thermal de Broglie wavelength of the radiation ($\propto T^{-1}$) is smaller than the Hubble radius ($H^{-1}$).

\vspace{0.2cm} 
The importance of studying warm inflation are manifold. First, it is natural and proper to include the inflaton couplings to  other fields not just in the reheating phase, but also during the  inflationary phase. 
Second, the condition of the flat potential required for the slow roll of inflaton is somewhat relaxed in warm inflation. As a result inflation can last long enough even if the potential is not very flat. Third, as pointed out in Refs. \cite{BasteroGil:2009ec,Panotopoulos:2015qwa,Visinelli:2016rhn,Benetti:2016jhf}, certain monomial potentials of inflation are viable models in warm inflation, unlike in cold inflation where they are ruled out.  Therefore it is crucial to reconsider the monomial models in warm inflation and constrain their parameters using the CMB observations. 

\vspace{0.2cm}
In our previous work \cite{Arya:2017zlb}, we obtained the model parameters for a $\lambda\phi^4$ potential model of warm inflation with a cubic dissipation coefficient. In this study, we consider a $\lambda \phi^4$ potential with a linear dissipation coefficient and a $\lambda \phi^6$ potential with linear and cubic dissipation coefficients and carry out a Markov Chain Monte Carlo (MCMC) analysis to obtain 
 the parameters consistent with the CMB observations. We parameterize the primordial power spectrum for our models in terms of two parameters, namely, $\lambda$, representing the inflaton self coupling,  and the dissipation parameter at the pivot scale, $Q_P$, due to the inflaton's interactions with other fields, and obtain their mean values. This is crucial for model building. We also verify that the $n_s$ and $r$ values  for these mean values are compatible with the observational data.
 
\vspace{0.2cm}
Warm inflationary models usually require coupling the inflaton to a very large number of fields to maintain $T>H$ \cite{BasteroGil:2009ec, Bastero-Gil:2015nja}. Such a large number of fields can be achieved through some string theory inspired generation mechanism \cite{Burgess:2001fx}.  However recently proposed warm inflation models with the inflaton as a pseudo-Nambu-Goldstone boson require very few additional fields \cite{Mishra:2011vh,Bastero-Gil:2016qru} and allow for a well-motivated particle physics description.

\vspace{0.2cm}
The objective of our work is to study scalar field monomial potentials in the context of warm inflation and estimate the parameters that are consistent with the observations using a publicly available numerical code named {\tt CosmoMC} \cite{cosmomc}. We use the \textit{ Planck} 2015 data (high-l TT, TE, TE + low-l polarization) for our analysis. We plot the joint probability distribution of our model parameters and list the marginalised values of all the parameters along with $68 \%$ C.L. values using the {\tt GetDist GUI} software.

\vspace{0.2cm}
This paper is organised as follows. We first overview the basic evolution equations for the inflaton and the radiation fields and define the slow roll parameters and conditions in Section \ref{theory}. We also state the conditions required for warm inflation to begin and to end. We describe the primordial power spectrum for warm inflation and the forms of the dissipation coefficient that we consider. In Section \ref{model}, we present the models that we consider. Then in Sections \ref{Model1}, \ref{Model2}, \ref{Model3}, we parameterize the primordial power spectrum for our models in terms of the model parameters, $\lambda$, and  $Q_P$.  
 In Section \ref{scale}, we determine the scale dependence of the power spectrum. Then in Section \ref{math}, we do a preliminary analysis using Mathematica and plot the dependences of the model parameters. After that, we carry out a 
 MCMC analysis using {\tt CosmoMC} and obtain the mean values for the parameters for the different models in Section \ref{result}. We also list the values of $n_s$ and $r$ for the mean values of the parameters. Lastly we end with our conclusions in Section \ref{conc}.

\vspace{0.2cm}
In our notation, overdot represents derivative w.r.t. time and prime represents derivative w.r.t. $\phi$ throughout this paper.
\\
\section{The theory of warm inflation}
\label{theory}

\subsection{Evolution equations for the inflaton and radiation }
The equation of motion of the homogeneous inflaton field $\phi$ during warm inflation is given as
\begin{equation}
\ddot \phi + (3 H + \Upsilon ) \dot\phi + V'(\phi)=0 \, ,
\label{inf}
\end{equation} 
where $H$ is the Hubble expansion rate, and $\Upsilon (\phi, T)$ is the dissipation coefficient which is a measure of inflaton dissipation into radiation. In the literature there are many forms of  $\Upsilon(\phi, T)$, which depend on the mechanism by which the inflaton dissipation takes place. In this study, we have considered temperature dependent forms of $\Upsilon$ which are described in Section \ref{upsilon}. The dissipative term $\Upsilon  \dot\phi$ is absent in the cold inflation scenario.

\vspace{0.2cm}
As a result of the inflaton dissipation, radiation is concurrently produced during warm inflation. From the continuity equation, the energy density of radiation $\rho_r$ evolves as
\begin{equation}
\dot\rho_r+4H\rho_r=\Upsilon{\dot\phi}^2~.
\label{rad}
\end{equation}
The energy dissipated by the inflaton, $\Upsilon{\dot\phi}^2$, is transferred to radiation. We define a dissipation parameter
$$Q \equiv \frac{\Upsilon}{3H} $$ and rewrite Eq. (\ref{inf}) as
\begin{equation}	
\ddot \phi + 3 H( 1+ Q ) \dot\phi + V'(\phi)=0.
\label{inflaton}
\end{equation}
The dissipation parameter, $Q$, is the ratio of the strength of inflaton dissipation into radiation to the Hubble rate of expansion.  For $Q\gg1$, the dissipation coefficient is larger than $H$, and this regime is termed as the strong dissipative regime. For $Q\ll 1$, the expansion is faster than dissipation, and this is termed as the weak dissipative regime of warm inflation. In this study, we have considered both the regimes for our models.
\\
\subsection{Slow roll parameters and conditions}
\label{slow}

The flatness of the potential $V(\phi)$ in warm inflation is measured in terms of 
the slow roll parameters which are defined in Ref. \cite{Hall:2003zp} as 
\begin{equation*}
\label{slow_roll_parameters}
\epsilon_\phi = 
\frac{M_{Pl}^2}{16\pi}\,\left(\frac{V_{,\phi}}{V}\right)^2, \hspace{1cm}
\eta_\phi = 
\frac{M_{Pl}^2}{8\pi}
\,\frac{V_{,\phi\phi}}{V},
\end{equation*}

\begin{equation}
 \hspace{-0.5cm}
\beta_\Upsilon = 
\frac{M_{Pl}^2}{8\pi}
\,\left(\frac{\Upsilon_{,\phi}\,V_{,\phi}}{\Upsilon\,V}\right),  \hspace{1cm}
\delta=\frac{T V_{,{\phi T}}}{V_{,\phi}}
\label{slowroll}
\end{equation}
Here $V_{,\phi}$ is the derivative of $V$ w.r.t. $\phi$ and $\Upsilon_{,\phi}$ is the derivative of $\Upsilon$ w.r.t. $\phi$. 
$M_{Pl}=\frac{1}{\sqrt{G_N}}=1.2\times 10^{19}$ GeV is the Planck mass.
In cold inflation, there is no $\beta_\Upsilon$ or $\delta$, as there is no dissipation and temperature during inflation. In the literature one also defines the horizon flow parameters as
\be
\epsilon_{H}=\frac{-\dot{H}}{H^2}, \qquad \quad  \eta_{H}=\frac{-\ddot{H}}{2H\dot H}.
\label{Horizonroll} 
\ee 
These different definition of slow roll parameters are then related as 
\be
\epsilon_{H}=\frac{\epsilon_{\phi}}{1+Q}, \qquad \eta_{H}=\frac{\eta_{\phi}}{1+Q}\,.
\label{horizon_slow}
\ee
The slow roll conditions needed for the warm inflationary phase are given in Ref. \cite{Moss:2008yb} as
\begin{equation} 
\label{slow_roll}
\epsilon_\phi \ll 1+Q,\quad |\eta_\phi| \ll 1+Q,\quad |\beta_\Upsilon| \ll 1+Q, \quad 0<\delta \ll \frac{Q}{1+Q}.
\end{equation}

\vspace{0.2cm}
We can see that for significant $Q$, these conditions reduce the requirement for the potential to be extremely flat. In the slow roll approximation, we can neglect $\ddot \phi$ in Eq. (\ref{inflaton}) which gives
\begin{equation}
\dot\phi\approx \frac{-V'(\phi)}{3H(1+Q)}.
\label{phido}
\end{equation}
\vspace{0.1cm}
\\
As shown in Fig. \ref{phiQkrho}, the energy density of radiation does not change appreciably when the modes of cosmological interest cross the horizon. Also, $\dot\rho_r$ is smaller than the other terms in Eq. (\ref{rad}) throughout inflation.  Therefore, we can approximate $\dot\rho_r\approx 0$ and obtain 
\begin{equation}
\rho_r=\frac{\pi^2}{30} g_* T^4\equiv A T^4\approx \frac{\Upsilon}{4H} {\dot\phi}^2 =\frac{3}{4} Q {\dot\phi}^2,
\label{rhodot}
\end{equation}
where $g_*$ is the number of relativistic degrees of freedom during warm inflation and we define $A=\pi^2g_*/30$. (Here we take $g_*\approx 200$.) 

\vspace{0.2cm}
We show the behaviour of the evolution of $\phi$ (in units of $M_{Pl}$) and the temperature $T$ during warm inflation for one of our models in Fig. \ref{phiQkrho}. The dissipation parameter, $Q$,  depends on both $\phi$ and $T$, and is not a constant but rather evolves during inflation. This behaviour can also be seen in Fig. \ref{phiQkrho}. It is possible that inflation starts in the weak dissipation regime, and with the evolution of $Q$, ends in the strong dissipation regime.

\vspace{0.2cm}
As already mentioned, the notion of a temperature 
is only valid for $T> H$. For the models we have considered, we find that $T>H$ at all times during inflation holds for $Q_P>10^{-5}$. 
 So we have a lower bound on the value of $Q_P$ of $10^{-5}$. 
\\
\subsection{End of warm inflation}
\label{end}

\begin{figure}[tbp]
\includegraphics[width=0.5\linewidth]{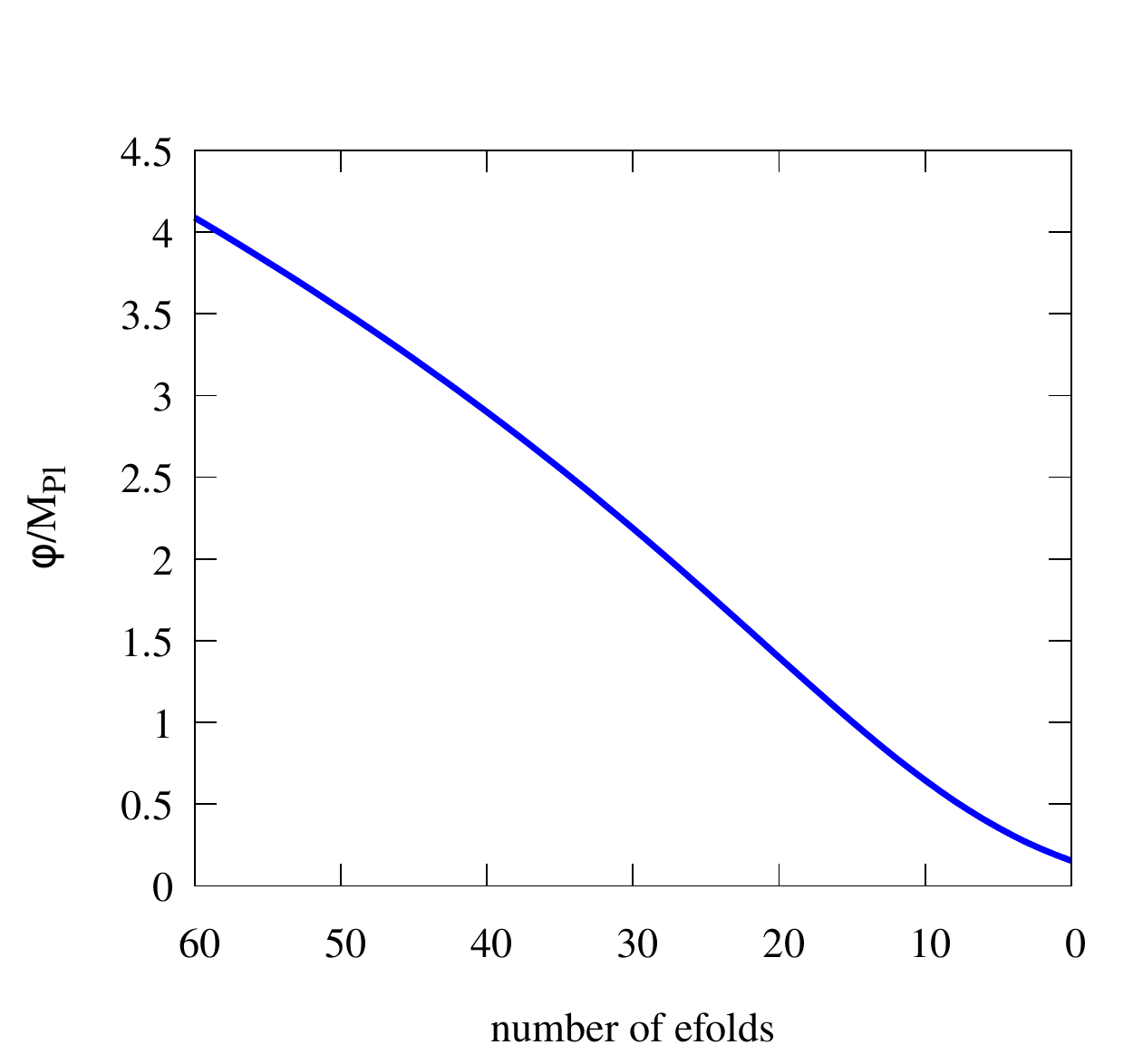}
\includegraphics[width=0.5\linewidth]{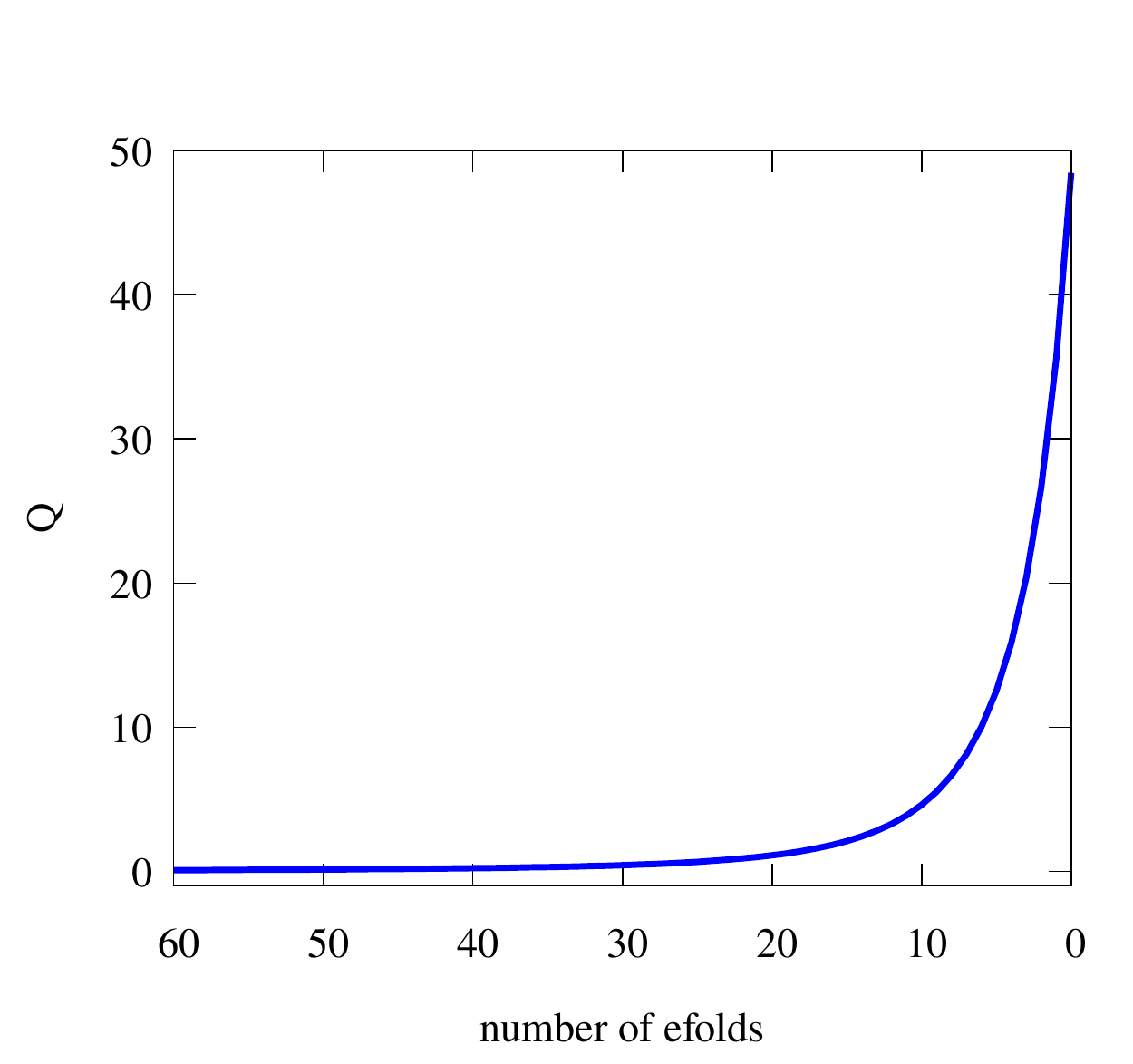}
\includegraphics[width=0.49\linewidth]{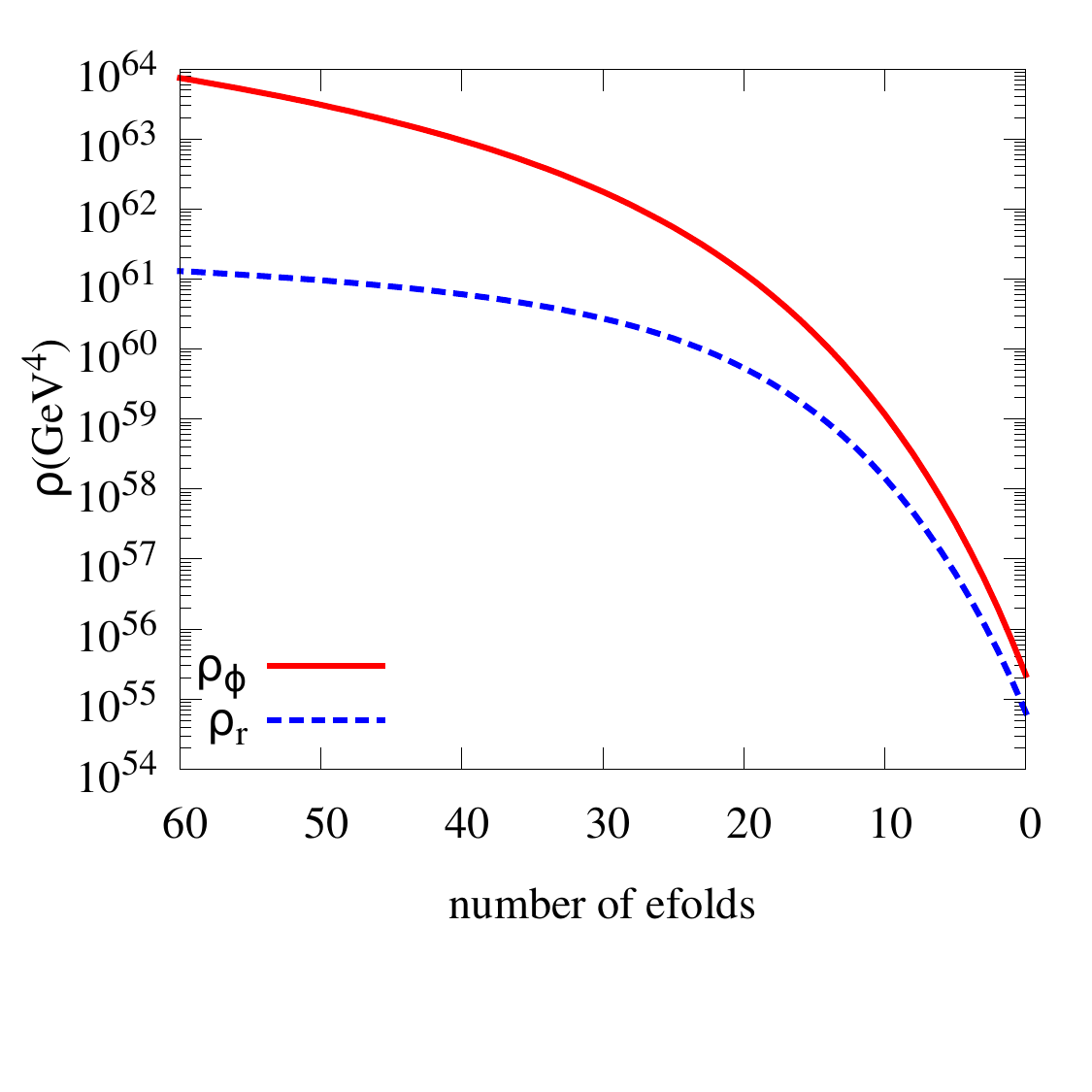}
\includegraphics[width=0.52\linewidth]{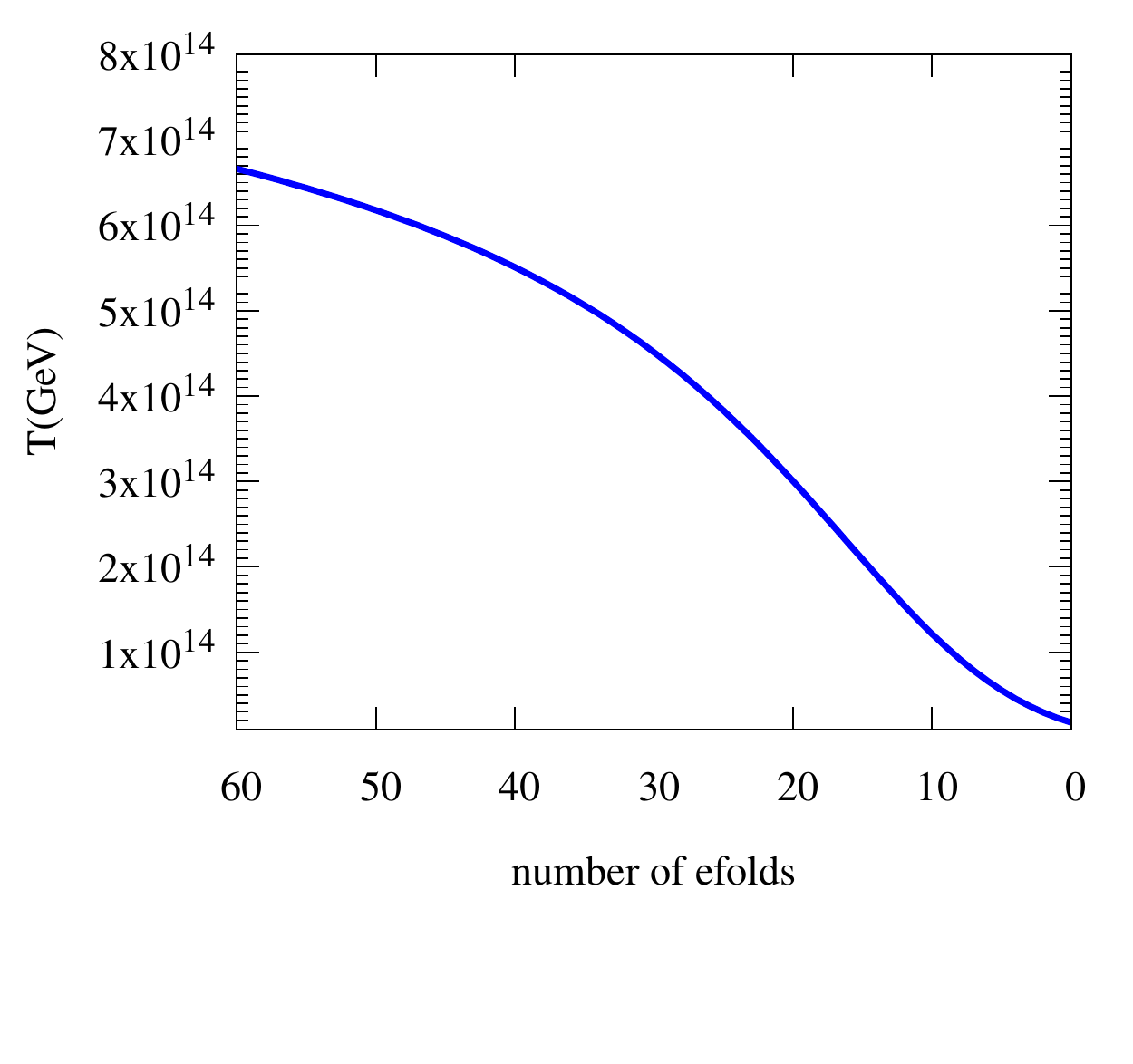}
\vspace{-1.5cm}
\caption{The behaviour of the homogeneous inflaton field $\phi$ (in units of $M_{Pl}$), the dissipation parameter $Q$, the energy density in $\phi$ (red solid) and radiation (blue dashed), and temperature $T$ of the Universe is shown as a function of the number of efolds $N$ counted from the end of inflation for  $V(\phi)=\lambda\frac{\phi^6}{M_{Pl}^2}  $ with the dissipation coefficient $\Upsilon= C_T T$. To generate this plot, we take 
$Q_P=0.1$, and as will be described later, fix the normalisation of the primordial
power spectrum at the pivot scale $P_\mathcal{R}(k_P)\equiv A_s=2.2\times10^{-9}$, and set $N_P=60$. This behaviour is similar for all the models we have considered. }
\label{phiQkrho}
\end{figure}

In the standard cold inflation, the violation of slow roll conditions marks the end of inflation. But warm inflation ends when either the slow roll conditions are violated or the radiation energy density dominates the inflaton energy density, i.e. $\rho_r>\rho_{\phi}.$  In Fig. \ref{phiQkrho}, we plot the radiation and inflaton energy density as a function of $N,$ the number of efolds from the end of inflation. For our models $\rho_\phi$ is larger than $\rho_r$, even when the slow roll conditions fail and  therefore the end of inflation is governed by the breaking of the slow roll conditions. As a consequence, the Universe will also go through a phase of reheating after inflation for the models we have considered.\\

\subsection{Different forms of the dissipation coefficient considered}
\label{upsilon}

In the literature, a general expression for the dissipation coefficient is given as
\begin{equation}\Upsilon(\phi, T) = C T^c \phi^{2a}/M_X^{2b}\end{equation}
with a condition $c+2a-2b=1$ \cite{Ramos:2013nsa}, where $M_X$ is the mass of the fields  $X$ which are coupled to the inflaton. Here we have considered the following two forms of the dissipation coefficient for the supersymmetric models of inflation 
in which the inflaton is coupled with the intermediate bosonic and fermionic components of a superfield, $ X$, which subsequently decay into the scalar and fermionic components of the superfield, $Y$  \cite{Moss}. The field $Y$  with $m_Y\ll T$ thermalise and constitute the thermal bath.

\begin{enumerate}
\item 
Dissipation coefficient with a cubic dependence on the temperature,
\begin{equation}
\Upsilon(\phi,T)=C_\phi \frac {T^3}{\phi^2}.
\label{upsiloncubic}
\end{equation}
This kind of dissipation term arises in the low temperature regime of warm inflation, where the intermediate $X$ fields are heavy ($m_X\gg T$) \cite{Moss,BasteroGil:2010pb,BasteroGil:2012cm}.
 The dimensionless constant $C_{\phi}$ depends on the couplings and the multiplicities of $X$ and $Y$ and is defined as 
$$C_\phi=\frac{1}{4}\alpha N_X $$
 where $\alpha=h^2 \frac{N_Y}{4\pi}$ is less than $1$, $h$ is the coupling between $\phi$ and $X$, and $N_{X, Y}$ are the multiplicities of the $X$ and $Y$ fields \cite{Imp}. $N_X$ can be very large; for example, for the sextic potential model that we consider 
 with a cubic dissipation coefficient, $C_\phi$ is $4\times 10^7$ for the mean values of the model parameters and so assuming $\alpha=0.1$, we get $N_X \sim 10^9$. Such a huge number of fields is a drawback for the warm inflationary models. 
 
 \vspace{0.2cm}
\item
Dissipation coefficient with a linear dependence on temperature, 
\begin{equation}
\Upsilon(\phi,T)=C_T T.
\label{linearupsilon}
\end{equation}
This kind of dissipation term arises in the high temperature regime of warm inflation,   where the $X$ fields are light, $T\gg m_X$ \cite{Moss,BasteroGil:2010pb}.\\
\end{enumerate} 

\subsection{Primordial power spectrum}
\label{power}
The primordial power spectrum for the warm inflation is given in Refs. \cite{Imp,Bastero-Gil:2016qru,Benetti:2016jhf,Ramos:2013nsa} (based on the references therein) as
\begin{equation}
P_\mathcal{R}(k)=\left(\frac{H_k^2}{2\pi\dot\phi_k}\right)^2\left[1+2n_k+\left(\frac{T_k}{H_k}\right)
\frac{2\sqrt{3}\pi Q_k}{\sqrt{3+4\pi Q_k}}\right]G(Q_k),
\label{power}
\end{equation}
where the subscript $k$ signifies the time when the  mode of cosmological perturbations with wavenumber $k$ leaves the horizon during inflation, and the inflaton distribution $n_k$ is taken as a Bose-Einstein distribution,
\begin{equation}
n_k=\frac{1}{\exp(\frac{k/a_k}{T_k}) -1}
\label{BE}
\nonumber
\end{equation}
where $a$ is the scale factor and $T$ represents the physical temperature. Then
\begin{equation}
1+2n_k= \coth\frac{H_k}{2T_k}.
\label{coth}
\nonumber
\end{equation}
The perturbations in the radiation, because of inhomogeneous dissipation, can lead to growing inflaton perturbations in the primordial power spectrum \cite{Graham:2009bf}. This growth factor $G(Q_k)$ is dependent on the form of $\Upsilon$ and is obtained numerically. As given in Ref. \cite{Bastero-Gil:2016qru,Benetti:2016jhf} 
$${\rm For\,\,} \Upsilon \propto T \qquad\qquad G(Q_k)_{linear}= 1+0.0185 \,Q_k^{2.315} +0.335 \,Q_k^{1.364}$$
$${\rm For\,\,} \Upsilon \propto T^3 \qquad\qquad G(Q_k)_{cubic}= 1+4.981 \,Q_k^{1.946} +0.127 \,Q_k^{4.330}.$$
In the weak dissipation regime for small $Q$, the growth factor does not enhance the power spectrum much. But for $Q>1$ in the strong dissipation regime, the growth factor significantly enhances the power spectrum.
 \\
 \vspace{0.1cm}
 \\
 The primordial tensor fluctuations of the metric give rise to a tensor power spectrum given in Ref. {\cite{Imp}} as
 \be
 P_T(k)= \frac{16}{\pi}\left(\frac{H_k}{M_{Pl}}\right)^2.
 \ee
The ratio of the tensor to the scalar power spectrum is expressed in terms of a parameter $r$ as
\be
r=\frac{P_T(k)}{P_\mathcal{R}(k)}.
 \ee
\section{Models of warm inflation considered} 
\label{model}
With the recent  
measurements of the temperature anisotropies in the Cosmic Microwave Background (CMB) radiation, tighter constraints on the spectral index of scalar perturbations, $n_s$ and the tensor-to-scalar ratio, $r$ are obtained which rule out simple monomial potential models of inflation like $V(\phi)= \lambda \phi^4$ and $V(\phi)= \lambda \phi^6$. In our earlier work, we obtained the parameters of warm inflation for 
a potential, $V(\phi)=\lambda \phi^4$, with a cubic dissipation coefficient $\Upsilon= C_{\phi} \frac{T^3}{\phi^2}$ in the weak dissipative regime \cite{Arya:2017zlb}. For the strong dissipative regime, the $n_s$ and $r$ values were out of the {\it Planck} 2015 allowed values and hence this case is not of interest. Here in this study, we have considered the following models in the weak and strong dissipative regimes.

\vspace{0.2cm}
\begin{itemize}
\item  
Model I: $V(\phi)=\lambda  \phi^4$ with the dissipation coefficient $\Upsilon= C_T T$. 
\item 
 Model II: $V(\phi)=\lambda\frac{\phi^6}{M_{Pl}^2}  $ with the dissipation coefficient $\Upsilon= C_T T$. 
\item
 Model III:
$V(\phi)=\lambda\frac{\phi^6}{M_{Pl}^2}  $ with the dissipation coefficient $\Upsilon= C_{\phi} \frac{T^3}{\phi^2}$. 
\end{itemize}

\vspace{0.2cm}
For all the models, we first parametrize the primordial power spectrum in terms of the model parameters. After a primary analysis of the models with  {\tt Mathematica}, we perform a detailed Markov Chain analysis with {\tt CosmoMC} and present the mean values of the parameters along with the $68\%$ confidence limits. Then for the mean values we compute $n_s$ and $r$ and compare their values with the {\textit {Planck}} 2015 results.

\vspace{0.2cm}	
 \section{Primordial power spectrum for $V(\phi)=\lambda  \phi^4$ with  $\Upsilon= C_T T$}
 \label{Model1}
 
 In this section, we evaluate each factor of the primordial power spectrum $P_{\mathcal R}$ given in Eq. (\ref{power}), and express them in terms of the  inflaton self coupling $\lambda$, dissipation parameter $Q$, and the dimensionless constant  $C_T$. 
  In Section \ref{scale}, we shall obtain $Q(k)$ which thereby gives $P_{\mathcal R}$ as a function of $k$, $\lambda$ and $C_T$. We will see in Section \ref{duration} that with another constraint on the number of efolds, the parameters $\lambda$ and $C_T$ are related. Then we will have only two independent model parameters, $\lambda$ and $Q_P$, for {\tt CosmoMC}{\footnote {This model was also considered in Ref. \cite{Bastero-Gil:2017wwl}, however, the power spectrum was parameterized differently.}}.\\
 
\begin{enumerate}
\item \textit{The prefactor:} The energy density during inflation is predominantly the potential of the inflaton field. Therefore we can write the Einstein equation as
\be H^2=\frac{8\pi}{3} \frac{\lambda \phi^4}{M_{Pl}^2}.
\label{Hubble1}
\ee
 Using this we can express Eq. (\ref{phido}) for this model as
\begin{equation}
\dot\phi\approx \frac{-V'(\phi)}{3H(1+Q)}= -\frac{4}{3} \sqrt{\frac{3}{8\pi}}\sqrt{\lambda}\frac{\phi M_{Pl}}{(1+Q)}.
\label{phidot1}
\end{equation}
Then combining these,  we can write {\footnote{The minus sign on the rhs was missed in Ref. \cite{Arya:2017zlb} but it does not affect the power spectrum.}}
\begin{equation}
\frac{H_k^2}{2\pi\dot\phi_k}= -
\sqrt{\frac{8\pi}{3}}\sqrt{\lambda}
\left(\frac{\phi_k}{M_{Pl}}\right)^3 (1+Q_k).
\label{front1}
\end{equation}
\vspace{0.01cm}
\item \textit{T/H factor:}
On substituting Eq. (\ref{phidot1}) in the energy density of radiation 
given in Eq. (\ref{rhodot}), we obtain the temperature of the thermal bath as
 \be
 T_k=\left(\frac{15}{\pi^3 g_*} \frac{Q_k}{(1+Q_k)^2} \lambda \,\phi_k^2 M_{Pl}^2\right)^\frac{1}{4}
 \label{T1}
 \ee
 and then combining with $H$ from Eq. (\ref{Hubble1}), the factor $T/H$ becomes
 \begin{equation}
 \frac{T_k}{H_k}=\left(\frac{15}{\pi^3 g_*}\right)^\frac{1}{4}\sqrt{\frac{3}{8\pi}}\lambda^{-\frac{1}{4}}
 \frac{Q_k^{\frac{1}{4}}}{(1+Q_k)^{\frac{1}{2}}} 	\left( \frac{\phi_k}{M_{Pl}}\right)^{-3/2}.
 \label{TkoverHk1}
 \end{equation}
 \end{enumerate}
 \vspace{0.2cm}
The dissipation parameter is defined as $Q=\frac{\Upsilon}{3H}$. In this model of warm inflation, we have considered $\Upsilon= C_T T.$ On substituting this form of $\Upsilon$ we get  
$T= \frac{3HQ}{C_T}.$
 We equate this with Eq. (\ref{T1}) and obtain
 \begin{equation}
 \left(\frac{\phi_k}{M_{Pl}}\right)=\sqrt\frac{1}{8\pi}\left(\frac{4 C_{T}^4}{ 9  \lambda A}\frac{1}{Q_k^3(1+Q_k)^2}\right)^{\frac{1}{6}}.
 \label{phikovermpl1}
 \end{equation}	
 \\
 \vspace{0.2cm}
 On substituting Eq. (\ref{phikovermpl1}) in Eqs. (\ref{front1}) and (\ref{TkoverHk1}), we can express $P_\mathcal{R}(k)$ in terms of variables $\lambda, Q_k$ and $C_T$. Also, from its definition in Eq. (\ref{Horizonroll}), the slow roll parameter can be written as
\be
\epsilon_{H}= 
\frac{8}{(1+Q_k)8\pi}\left(\frac{M_{Pl}}{\phi_k}\right)^2=\frac{8}{1+Q_k} 
\left(\frac{9 A \lambda}{4 C_T^4}{Q_k^3(1+Q_k)^2}\right)^\frac{1}{3}.
\label{eph1}
\ee
It can also be seen that the slow roll parameters are related as
$$\epsilon_{H}=\frac{2}{3}\eta_{H} 
\qquad {\rm and}\quad \beta_\Upsilon=0.$$
Using Eq. ({\ref{Hubble1}}), the tensor power spectrum for this model is evaluated as
\be
P_T(k)= \frac{16}{\pi}\left(\frac{H_k}{M_{Pl}}\right)^2=\frac{128}{3}\lambda \left(\frac{\phi}{M_{Pl}}\right)^4.
\ee
Further, we can use Eq. (\ref{phikovermpl1}) and express $P_T(k)$ in terms of model parameters.
\\
 \section{Primordial power spectrum for $V(\phi)=\lambda\frac{\phi^6}{M_{Pl}^2}$ with $\Upsilon= C_T T$}
 \label{Model2}
 In this model also we rewrite the primordial power spectrum in terms of $\lambda$, $Q$, and $C_T$ and similarly we will see that with another constraint on the number of efolds, the parameters $\lambda$ and  $C_T$ are related. Then we will have only two independent model parameters, $\lambda$ and $Q_P$ for the {\tt CosmoMC} analysis.\\
 \begin{enumerate}
 \item \textit{The prefactor:} For this inflaton potential, 
 the Einstein equation is given as
 \be 
 H^2=\frac{8\pi}{3} \frac{\lambda \phi^6}{M_{Pl}^4}.
 \label{Hubble2}
 \ee
 Using this we can express Eq. (\ref{phido}) for this model as 
\begin{equation}
\dot\phi\approx \frac{-V'(\phi)}{3H(1+Q)}= -2 \sqrt{\frac{3}{8\pi}}\sqrt{\lambda}\frac{\phi^2}{(1+Q)}.
\label{phidot2}
\end{equation}
Then on combining these, we can write 
\begin{equation}
\frac{H_k^2}{2\pi\dot\phi_k}=-\frac{2}{3}
\sqrt{\frac{8\pi}{3}}\sqrt{\lambda}
 \left(\frac{\phi_k}{M_{Pl}}\right)^4  (1+Q_k).
 \label{front2}
 \end{equation}
 \vspace{0.01cm}
 \item \textit{T/H factor:}
On substituting Eq. (\ref{phidot2}) in the energy density of radiation 
given in Eq. (\ref{rhodot}), we obtain the temperature of the thermal bath as
 \be
T_k=\left(\frac{135}{4\pi^3 g_*} \frac{Q_k}{(1+Q_k)^2} \lambda \,\left(\frac{\phi_k}{M_{Pl}}\right)^4\right)^\frac{1}{4} M_{Pl}.
 \label{T2}\vspace{-0.1cm}
 \ee
and then using Eq. (\ref{Hubble2}), the factor $T/H$ becomes
\begin{equation}
\frac{T_k}{H_k}=\left(\frac{135}{4\pi^3 g_*}\right)^\frac{1}{4}\sqrt{\frac{3}{8\pi}}\lambda^{-\frac{1}{4}}
\frac{Q_k^{\frac{1}{4}}}{(1+Q_k)^{\frac{1}{2}}} 
\left( \frac{\phi_k}{M_{Pl}}\right)^{-2}.
 \label{TkoverHk2}
\end{equation}
\end{enumerate}
\vspace{0.2cm}
In this model of warm inflation, we have considered $\Upsilon= C_T T$. On substituting this form in the definition of $Q$, we get  
$T= \frac{3HQ}{C_T}.
\label{T} $
We equate this with Eq. (\ref{T2}) and obtain
\begin{equation}
 \left(\frac{\phi_k}{M_{Pl}}\right)=\left(\frac{ C_{T}^4}{ \lambda A}\frac{1}{(8 \pi)^3 Q_k^3(1+Q_k)^2}\right)^{\frac{1}{8}}~.
 \label{phikovermpl22}
 \end{equation}
 \\
 \vspace{0.2cm}
 On substituting Eq. (\ref{phikovermpl22}) in Eqs. (\ref{front2}) and (\ref{TkoverHk2}), we can express $P_\mathcal{R}(k)$ in terms of variables $\lambda, Q_k$ and $C_T$. Also, from its definition, the slow roll parameter can be written as	
 \be
 \epsilon_{H}=\frac{\epsilon_{\phi}}{1+Q_k}=\frac{18}{8\pi(1+Q_k)}\left(\frac{M_{Pl}}{\phi_k}\right)^2=\frac{18}{1+Q_k}
 \left(\frac{A \lambda}{ 8\pi C_{T}^4}{Q_k^3(1+Q_k)^2}\right)^\frac{1}{4}
 \label{eph2}
 \ee
For this model, it can be seen that the different slow roll parameters are related as
 $$\epsilon_{H}=\frac{3}{5}\eta_{H} \quad {\rm and} \quad \beta_\Upsilon=0. $$
Using Eq. ({\ref{Hubble2}}), the tensor power spectrum for this model is evaluated as
\be
P_T(k)= \frac{16}{\pi}\left(\frac{H_k}{M_{Pl}}\right)^2=\frac{128}{3}\lambda \left(\frac{\phi}{M_{Pl}}\right)^6.
\label{tensor2}
\ee
Further, we can use Eq. (\ref{phikovermpl22}) and express $P_T(k)$ in terms of model parameters.
\\
\section{Primordial power spectrum for $V(\phi)=\lambda  \frac{\phi^6}{M_{Pl}^2}$ with  $\Upsilon=C_\phi \frac{T^3}{\phi^2} $}
 \label{Model3}
The inflaton potential $V(\phi)$ in this case is the same as in the previous model. Therefore the \textit{prefactor} and the \textit{T/H factor} for this model are the same as in Section \ref{Model2}. In this warm inflation model, we have considered $\Upsilon= C_\phi \frac{T^3}{\phi^2}$.
From the definition of $Q$, we get
$T=\left(\frac{3Q\phi^2 H}{C_{\phi}}\right)^{1/3}.
\label{T3} $
On equating this with Eq. (\ref{T2}), we obtain
\begin{equation}
\left(\frac{\phi_k}{M_{Pl}}\right)=\sqrt\frac{1}{8\pi}\left(\frac{81 \lambda C_{\phi}^4}{ 8 \pi A^3}\frac{1}{Q_k(1+Q_k)^6}\right)^{\frac{1}{8}}~.
\label{phikovermpl3}
\end{equation}	
\vspace{0.2cm}
 Using this, the slow roll parameter is expressed as
 \be
 \epsilon_{H}=\frac{\epsilon_{\phi}}{1+Q_k}=\frac{18}{8\pi(1+Q_k)}\left(\frac{M_{Pl}}{\phi_k}\right)^2=\frac{18}{1+Q_k}
 \left(\frac{ 8 \pi A^3}{ 81 \lambda C_{\phi}^4}Q_k(1+Q_k)^6\right)^\frac{1}{4}.
 \label{eph3}
 \ee
 We can see that in this case the different slow roll parameters are related as
 $$\epsilon_{H}=\frac{3}{5} \, \eta_{H}
 =\frac{1}{(1+Q)}\frac{3}{2} \, |\beta_\Upsilon|.
 $$  
 The tensor power spectrum for this model is the same as given in Eq. (\ref{tensor2}) for Model II.
 \\
 \section{Scale dependence of the power spectrum, $P_\mathcal{R}(k)$}
 \label{scale}
 
 After effectively obtainig the primordial power spectrum for all the models in terms of $Q,\lambda$, and $C_\phi$ or $C_T$, we now proceed to obtain $Q(k)$ and thus $P_\mathcal{R}(k)$. Firstly we write 
 \be
 \frac{dQ_k}{dx}=\frac{dQ}{dN}\frac{dN_k}{dx}\, ,
 \label{dQbydx}
 \ee 
 where we have defined a variable $x={\rm ln} \frac{k}{k_P}$, and $k_P$ corresponds to the pivot scale. Then integrating $\frac{dQ_k}{dx}$ gives us $Q_k(k)$. This is carried out as below.\\
\vspace{0.2cm}
\subsection{Obtaining $\frac{dQ}{dN}$}

\label{Qk}
In this section we will evaluate how the dissipation parameter, $Q$, evolves with the number of efolds, $N$. 
\vspace{0.2cm}
\begin{enumerate}
 \item 
 For \textit{Model I}, we first differentiate Eq. (\ref{phikovermpl1}) w.r.t $N$ and write  $\frac{d \phi}{dN}=\frac{d \phi}{dt}\frac{d t}{dN}=\frac{ \dot\phi}{-H}.
 $ Then by using Eqs. (\ref{Hubble1}), (\ref{phidot1}) and  (\ref{phikovermpl1}), we obtain
 \begin{equation} 
 \frac{dQ}{dN}= 
 -24\left(\frac{9 A \lambda}{4C_{T}^4}\right)^{\frac{1}{3}}\frac{Q^{2}(1+Q)^{2/3}}{3+5Q}.
 \label{dQdN1}
 \end{equation}
 \item 
 Similarly for \textit{Model II}, we differentiate Eq. (\ref{phikovermpl22}) w.r.t $N$ and then again write $\frac{d \phi}{dN}=\frac{ \dot\phi}{-H}.$ 
 By using Eqs. (\ref{Hubble2}), (\ref{phidot2}) and (\ref{phikovermpl22}), we obtain
 \begin{equation} 
 \frac{dQ}{dN}= 
 -\frac{6}{\pi}\left(\frac{A \lambda (8\pi)^3}{C_{T}^4}\right)^{\frac{1}{4}}\frac{Q^{7/4}(1+Q)^{1/2}}{3+5Q}~.
 \label{dQdN2}
 \end{equation}
\item 
 For \textit{Model III}, on differentiating Eq. (\ref{phikovermpl3}) w.r.t $N$,
 and then using Eqs. (\ref{Hubble2}), (\ref{phidot2}) and (\ref{phikovermpl3}), we obtain
 \begin{equation} 
 \frac{dQ}{dN}= 
 -16\left(
 \frac{8 \pi A^3}{\lambda C_{\phi}^4}\right)^{\frac{1}{4}}\frac{Q^{5/4}(1+Q)^{3/2}}{1+7Q}~.
 \label{dQdN3}
 \end{equation}
 \end{enumerate}

 \subsection{Obtaining $\frac{dN}{dx}$}
 \label{number}
 \vspace{0.2cm}
 First we define number of efolds at any scale $k$, $N_k$, as 
 $$N_k= \ln \frac{a_e}{a_k}=\ln \frac{a_e}{a_P}+\ln\frac{a_P}{a_k}$$
 where $a_e, a_P,$ and $a_k$ are the scale factor at the end of inflation, and at the time when the pivot scale and the $k^{th}$ scale cross the horizon respectively. We also have $k=aH$. This gives 
 \be
 N_k = N_P+ \ln\frac{k_P H_k}{k H_P}
 =  N_P-\ln\frac{k}{k_P}+\ln\frac{H_k}{H_P}.
 \label{N1}
 \ee
 On differentiating the above equation w.r.t $x$, we obtain  
 \begin{equation}
 \frac{dN_k}{dx}=-1+\frac{\dot H_k}{H_k}\frac{dt}{dN}\frac{dN_k}{dx}=-1-\frac{\dot H_k}{H_k^2}\frac{dN_k}{dx}. 
 \label{dNd}
 \end{equation}
 Now from the definition of $\epsilon_H=-\frac{\dot H}{H^2}$, we can write Eq. (\ref{dNd}) as
 \begin{equation}
 \frac{dN_k}{dx}=-\frac{1}{1-\epsilon_H} 
 \label{dNdx}\,.
 \end{equation}
 \\
 \vspace{0.2cm}
 Now we combine Eqs (\ref{dQdN1}), (\ref{dQdN2}), (\ref{dQdN3}) with Eq. (\ref{dNdx}) and obtain $\frac{dQ_k}{dx}$ for all the models. The expressions obtained can then be integrated in {\tt Mathematica} from $Q_P$ (where $x=0)$ to any $Q_k$ (where $x={\rm ln} \frac{k}{k_P}$) to obtain $Q_k(k).$
 \vspace{0.2cm}
\\
 \subsection{Relation between $\lambda$ and $C_{\phi}$ or $C_T$}
 \label{duration}
  \vspace{0.2cm}
 In this section we will show that for a given $Q_P$, if we fix the  number of efolds for the pivot scale, $N_P$, then the parameters $\lambda$ and $C_\phi$ or $C_T$ are related.

 \vspace{0.2cm}
 As mentioned before, inflation ends when the slow roll conditions do not hold or when the energy density of the radiation begins to dominate. In our models we find that the failure of slow roll conditions decides the end of warm inflation. We see that $\eta_\phi$ is the largest amongst all the slow roll parameters. Hence the breakdown of $\eta_\phi<1+Q$ condition will decide the end of inflation.\\
 
 \begin{enumerate}
 	\item
 	 For $V(\phi)=\lambda\phi^4$, and for a linear dissipation coefficient, we obtain $Q_e$ at the end of inflation by setting the slow roll parameter
 	\be
 	\eta_e=\frac{12}{8\pi}\frac{M_{Pl}^2}{\phi_e^2}=1+Q_e.	
 	\ee
 On substituting Eq. (\ref{phikovermpl1}), we get the equation for $Q_e$ as
 	\be
 	Q_e^3(1+ Q_e)^{-1} = \frac{4C_{T}^4}{9 A \lambda}\left(\frac{1}{12}\right)^3.
 	\label{qe1}
 	\ee
 We solve this equation in {\tt Mathematica} and obtain $Q_e$ as some function of $\lambda$ and $C_T$. \\
 \vspace{0.1cm}
 
 	\item 
 	For $V(\phi)=\lambda\frac{\phi^6}{M_{Pl}^2}$,
 	by setting 
 	\vspace{-0.2cm}
 	\be
 	\eta_e=\frac{30}{8\pi}\frac{M_{Pl}^2}{\phi_e^2}=1+Q_e,
 	\label{eta}
 	\ee
 	and on substituting Eq. (\ref{phikovermpl22}), we obtain the equation for a linear dissipation coefficient as
 	\be
 	Q_e^3(1+ Q_e)^{-2} = \frac{C_{T}^4 8 \pi}{A \lambda}\left(\frac{1}{30}\right)^4.
 	\label{qe2}
 	\ee
 	On substituting Eq. (\ref{phikovermpl3}), we obtain the equation for a cubic dissipation coefficient as
 	\be
 	Q_e^3+2Q_e^2+Q_e = \frac{\lambda C_{\phi}^4 }{8 \pi A^3}\left(\frac{1}{10}\right)^4.
 	\label{qe3}
 	\ee	
 	On solving these equations in {\tt Mathematica}, we can obtain $Q_e$ as some function of $\lambda$ and $C_\phi$ (cubic dissipation) or $\lambda$ and $C_T$ (linear dissipation).
 \end{enumerate}
 \vspace{0.3cm}
 The solutions for $Q_e$ for all the models are given in Appendix \ref{sol}.
 Next we integrate $dN/dQ=(dQ/dN)^{-1}$ 
 from $Q_P$ (where $N=N_P$) to $Q_e$ (where $N=N_e=0$)
 using the expressions for ${dQ}/{dN}$ obtained in Section \ref{Qk} for all the models.
 We get $N_P$ as some function of $Q_P$ and $Q_e$ (which itself depends on $\lambda$ and $C_\phi$ or $C_T$, as seen above). 

\be N_P=F(Q_e)-F(Q_P)= F(\lambda,C_\phi \,{\rm or}\, C_T)-F(Q_P) \, ,
\label{Np}
\ee
\\
 where the integral function $F(Q)$ is given in Appendix \ref{sol} for all the models. This implies that for a given $Q_P$, if we fix $N_P$, our $\lambda$ and $C_\phi$ or $C_T$ will be related through Eq. (\ref{Np}). We employ this procedure to effectively obtain the power spectrum as a function of only two parameters, $\lambda$ and $Q_P$ after fixing $N_P$. 
\\
\subsection{Spectral index}
\label{spectral}
 \vspace{0.2cm}
The \textbf{spectral index} of the primordial power spectrum is defined as
\be
n_s-1= \left.\frac{d\ln P_\mathcal{R}(k)}{d\ln(k/k_P)}\right\rvert_{k=k_P}=\left.\frac{d\ln P_\mathcal{R}}{d Q}\frac{dQ}{dN} \frac{dN}{dx}\right\rvert_{k=k_P}.
\nonumber
\label{nsfull}
\ee 	
It measures the tilt of the power spectrum at the pivot scale ($k_P=0.05$ Mpc$^{-1}$).  The expressions for the spectral index for all the models are given in Appendix \ref{ns}. \\

\begin{figure}[h]
     \subfigure[]{\label{QL}
     \includegraphics[width=0.43\linewidth]{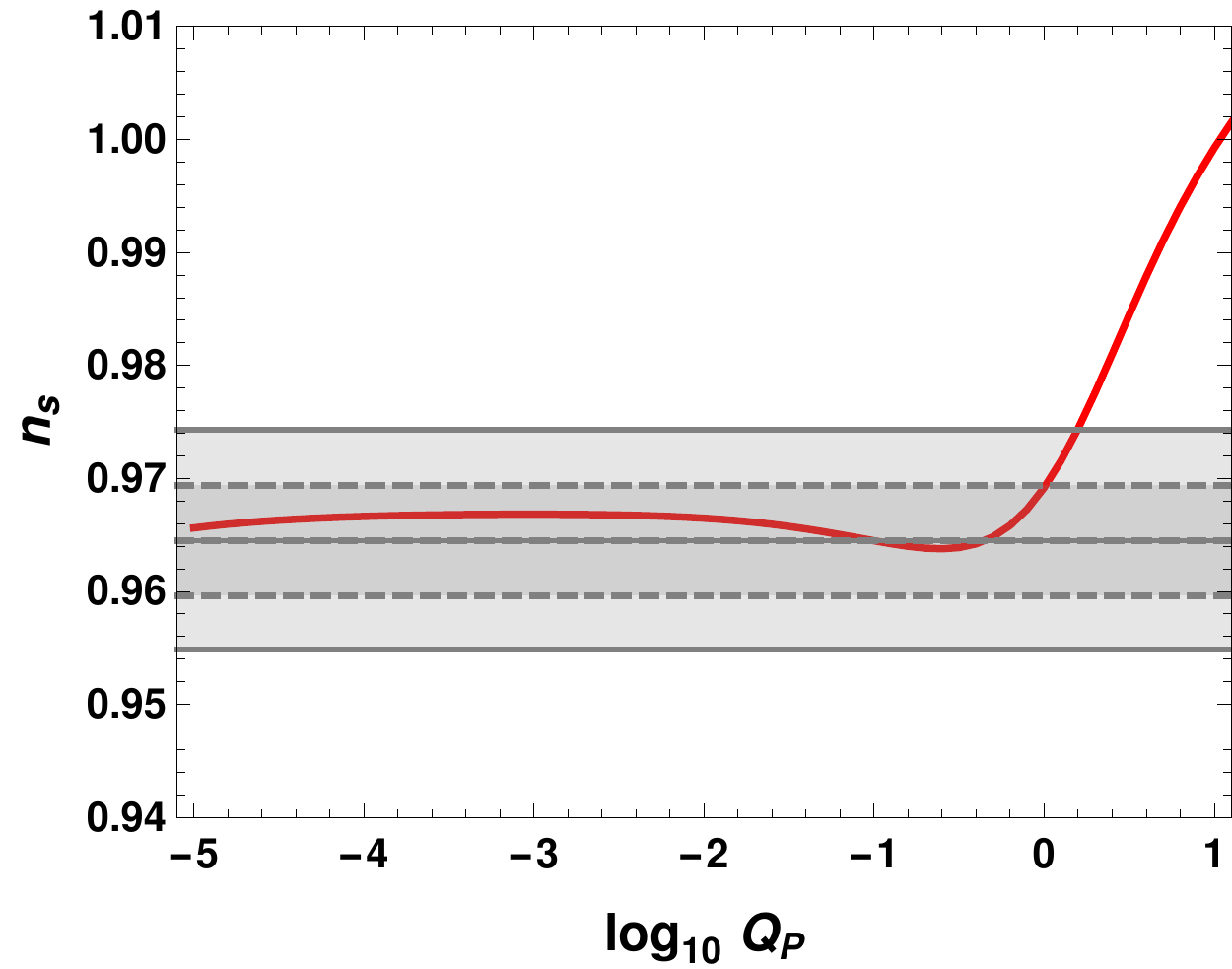}}
     \subfigure[]{\label{SL}
     \includegraphics[width=0.43\linewidth]{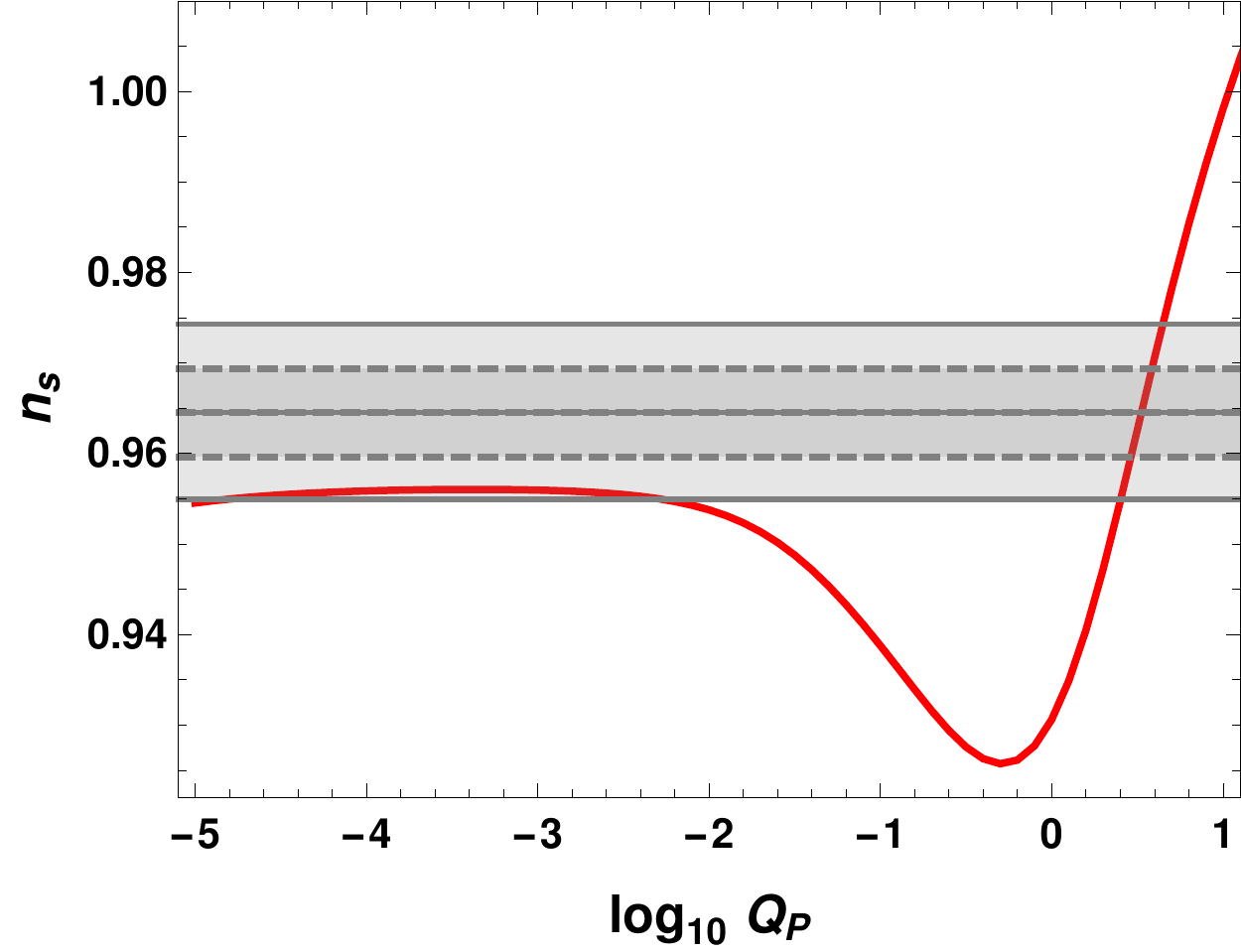}}
     \vspace{-0.1cm}
     \begin{center}
     \subfigure[]{\label{SC}
     \includegraphics[width=0.43\linewidth]{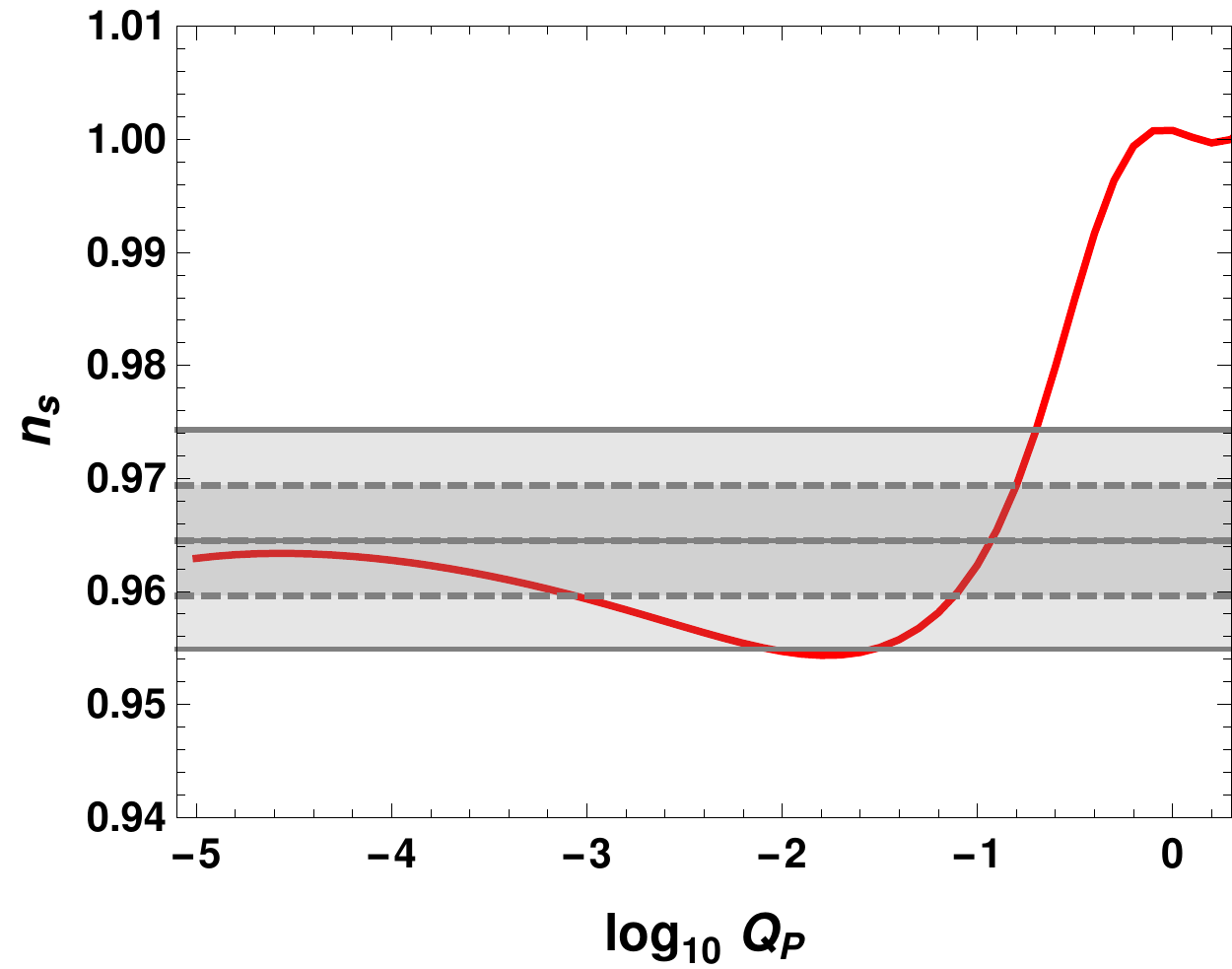}}
     \end{center}
     \vspace{-0.5cm}
     \caption{The spectral index, $n_s$ as a function of the dissipation parameter, $\log_{10} Q_P$ are plotted here (keeping $P_\mathcal{R}(k_P)=A_s=2.2\times10^{-9}$ and $N_P=60$) for  $V(\phi)=\lambda  \phi^4$ with linear dissipation (Fig. \ref{QL})
     and $V(\phi)=\lambda  \frac{\phi^6}{M_{Pl}^2}$ with linear and cubic dissipation (Fig. \ref{SL} and \ref{SC} respectively). The band shows the allowed $n_s$ values from \textit{Planck} 2015 results for a power law power spectrum with 68$\%$ 
     and 95$\%$ C.L..  Each point on the plots corresponds to the values of $\lambda$ and $C_\phi$ or $C_T$ consistent
     with the power spectrum normalisation and values of $Q_P$ and $N_P$.} 
     \label{fig:ns}
     \end{figure}

\vspace{0.2cm}
 The behaviour of the spectral index $n_s$ as a function of the dissipation parameter $Q_P$ for all the models we have considered is shown in Fig. \ref{fig:ns}. To generate these plots in {\tt Mathematica}, we fixed $P_\mathcal{R}(k_P)=A_s=2.2\times10^{-9}$ (the value is taken from Ref. \cite{Ade:2015xua}) and $N_P=60$. For a given $Q_P$ and a fixed $N_P$, when we fix the normalisation, $A_s$,  
 we obtain $\lambda$ and $C_\phi$ or $C_T$.  We then obtain $n_s$.
 The colored bands show $68 \%$ and $95\%$ C.L. allowed values for $n_s$ for a power law power spectrum from the \textit{Planck} 2015 TT, TE, EE+low P results \cite{Ade:2015xua,Ade:2015lrj,wiki}.
 \\ 
   
   \vspace{0.2cm}
   For $V(\phi)=\lambda\phi^4$ and $\lambda\frac{\phi^6}{M_{Pl}^2}$ with the linear dissipation coefficient $\Upsilon=C_T T$,  we have considered both weak and strong dissipative regimes, as shown in Fig. \ref{QL} and \ref{SL}. We see from Fig. \ref{SL} that in the weak dissipative regime, not all the values of $Q_P$ are consistent with observations. Hence for our {\tt CosmoMC} analysis we run over that particular range of $Q_P$ only for which $n_s$ is within the allowed range. We also find from Fig. \ref{SC} for  $V(\phi)=\lambda\frac{\phi^6}{M_{Pl}^2}$  with  the cubic dissipation coefficient $\Upsilon=C_\phi \frac{T^3}{\phi^2}$, that none of the values of $Q_P$ in the strong dissipative regime are consistent with the allowed range of values. Hence, we consider only the weak dissipative regime for running {\tt CosmoMC} for this model.  
  \\
 \section{Preliminary analysis of the parameters with Mathematica}
 \label{math}

In our models, the primordial power spectrum is written in terms of only two model parameters, $\lambda$ and $Q_P$. For our primary analysis in Mathematica, we use the normalisation condition $A_s=2.2\times 10^{-9}$ and obtain the values for $\lambda$ 
 for different $Q_P$. However in {\tt CosmoMC}, we run $\lambda$ and $Q_P$ independently over an estimated range without the normalisation constraint.
 We generate the $\lambda-Q_P$ plots for the weak and strong dissipative regimes of all the models as shown in the \textit{Left} plots of Figs.  \ref{fig:QL}, \ref{fig:SL}, and \ref{fig:SC}. We expect a similar behaviour to be observed in the {\tt CosmoMC} analysis. We find that the $\lambda-Q_P$ behaviour for the weak dissipative regime is different from the strong dissipative regime by comparing the slopes of the $\log \lambda-\log Q_P$ plots.  We also estimate the $\lambda-Q_P$ relation in Section \ref{result}.
\vspace{0.2cm}
 \section{CosmoMC results}
 \label{result}
 We have carried out a Markov Chain Monte Carlo (MCMC) analysis of our models for the weak and strong dissipation regime using a numerical code {\tt CosmoMC} (Cosmological Monte Carlo).  We run the {\tt CosmoMC} chains run over a six dimensional parameter space with flat priors on the baryon density $\Omega_bh^2$, cold dark matter density $\Omega_ch^2$, the observed angular size of the sound horizon at recombination $ \theta$, and the reionization optical depth $\tau$.
 Along with these, we choose $-\log_{10} \lambda$ and $-\log_{10} Q_P$ as independent model variables for the weak dissipative regime and  $-\log_{10} \lambda$ and $\log_{10} Q_P$ as independent model variables for the strong dissipative regime in performing our analysis. We take $N_P=60$ for our entire analysis.\\
 
\vspace{0.1cm}
 We use the September 2017 version of {\tt CAMB}
 and the November 2016 version of {\tt CosmoMC} and set the flags,  compute\_tensor=T, CMB\_lensing=F, and use\_nonlinear\_lensing=F. We use as the pivot scale $k_{P}=0.05 \, \text{Mpc}^{-1}$, and perform the CosmoMC analysis with the  \textit{Planck} 2015 TT, TE, EE + low P dataset.\\
 
 \vspace{0.1cm}
 We give below the priors for the parameters of our models and the mean values, along with 68\% confidence limits, obtained in the weak and the strong dissipative regimes. 
  For the mean values of our model parameters, $\lambda$ and $Q_P$, we also estimate the $n_s$ and $r$. The allowed values of $n_s$ and $r$ from the \textit{Planck} 2015 results (TT, TE, EE+ lowP) \cite{Ade:2015xua,Ade:2015lrj,wiki} are 
  \begin{equation*}
  n_s=0.9645 \pm 0.0049\, (68\%\rm \, C.L.) \qquad and \qquad
  0.9645^{+0.0098}_{-0.0096} \,(95\%\rm \,C.L.)
  \end{equation*}
  \begin{equation}
   r_{0.002}<0.10 \,(95\% \,\rm {C.L.}). 
   \label{constraint}
  \end{equation} 
  \\
 As mentioned in Ref. {\cite{Ade:2015xua}}, {\it Planck} is sensitive to the tensor modes in the low-l temperature power spectrum and hence
 the value of the tensor-to-scalar ratio is given at the scale of $0.002$ Mpc$^{-1}$. 
 However, it can be related to the tensor-to-scalar ratio at the scale of
 $0.05$ Mpc$^{-1}$ with the relation given in footnote 3 of Ref. \cite{Ade:2015xua}, and is not very different.
 \\
 \subsection{Model I: $V(\phi)=\lambda  \phi^4$ and $\Upsilon= C_T T$}
 We write the priors for the parameters of our model and the mean values along with 68\% confidence limits in the weak and the strong dissipative regimes in Tables \ref{tab:wQL} and \ref{tab:sQL}.
 \\
 \begin{table}[h!]
 	\begin{minipage}{0.5\textwidth}
\begin{tabular}{c c c} \hline\hline
 { Parameter} & { Priors}  & 68\% limits \vspace{0.1cm} \\ \hline \vspace{0.1cm}
$\Omega_bh^2$ & [0.005,0.1]  & $   0.02168\pm 0.00014  $ 
 	\vspace{0.1cm}
 	\\
 $\Omega_ch^2$ & [0.001,0.99] & $   0.1217\pm 0.0010$ 
 	\vspace{0.1cm}
 	\\
 $100\theta$   & [0.50,10.0]  &  $  1.04027^{+0.00029 }_{-0.00033}$\vspace{0.1cm}
 	\\ 
 $\tau$ &  [0.01,0.8]   & $ 0.048 ^{+0.016 }_{-0.031}$  \vspace{0.1cm}
 	\\ 
 $-\log_{10}\lambda$ & [13.7,15.5]& $ 14.39^{+0.34}_{-0.24}$\vspace{0.1cm}
 	\\ 
 $-\log_{10}Q_P$ &  [0.0,5.4]  & $  3.64 ^{+0.76}_{-1.1}$\vspace{0.1cm}
 	\\ 
 \hline\hline
 \end{tabular}
 		
\end{minipage}\hfill
 \begin{minipage}{0.4\textwidth}
Mean value of $\lambda$ = $4.07 \times 10^{-15}$\\
 Mean value of $Q_P$ = $2.29 \times 10^{-4}$
 \vspace{0.3cm}
 \\
 For these values, we obtain\\ 
 $n_s=0.967$\\
 $r=0.0330$
\end{minipage}
\caption{The priors and the marginalised values along with 68\% limits for the parameters of the model  $V(\phi)=\lambda  \phi^4$ with  $\Upsilon= C_T T$ in the weak dissipative regime are shown here. The mean values of the model parameters and the corresponding values of $n_s$ and $r$ are also given.  
}
\label{tab:wQL}
 \end{table}

  \begin{table}[h!] \vspace{0.5cm}
   \begin{minipage}{0.5\textwidth}
    \begin{tabular}{c c c} \hline\hline
   { Parameter} & { Priors}  & 68\% limits \vspace{0.1cm} \\ \hline \vspace{0.1cm}
   $\Omega_bh^2$ & [0.005,0.1]  & $   0.02174\pm 0.00013  $ 
   \vspace{0.1cm}
   \\
   $\Omega_ch^2$ & [0.001,0.99] & $   0.1200\pm0.0011$ 
   \vspace{0.1cm}
   \\
    $100\theta$   & [0.50,10.0]  &  $  1.04044\pm0.00029 $\vspace{0.1cm}
    	\\ 
    $\tau$ &  [0.01,0.8]   & $ 0.061\pm0.024$  \vspace{0.1cm}
    	\\ 
    $-\log_{10}\lambda$ & [15.0,15.6]& $ 15.166^{+0.036 }_{-0.056}$\vspace{0.1cm}
    	\\ 
    $\log_{10}Q_P$ &  [0.0,0.6]  & $ <0.156  $\vspace{0.1cm}
    \\ 
    \hline\hline
    \end{tabular}
    \end{minipage}\hfill
    \begin{minipage}{0.4\textwidth}
   Mean value of $\lambda$ = $6.82\times 10^{-16}$\\
   Upper limit of $Q_P$ $=1.43$
    \vspace{0.3cm}
    \\
   For the upper limit, we obtain\\ 
    $n_s=0.973$\\
    $r=0.000214$
   \end{minipage}
   \caption{The priors and the marginalised values along with 68\% limits for the parameters of the model  $V(\phi)=\lambda  \phi^4$ with  $\Upsilon= C_T T$ in the strong dissipative regime are shown here. The mean values of the model parameters and the corresponding values of $n_s$ and $r$ are also given. 
    }
   \label{tab:sQL}
   \end{table}
   \vspace{0.2cm}
  We can see that the mean values of $Q_P$ obtained from {\tt CosmoMC} for the weak and strong dissipative regimes lie in the allowed range of values of $Q_P$ in the $n_s$ plot in Fig. \ref{QL}.  The values of $n_s$ and $r$ obtained from the mean values of $\lambda$ and $Q_P$ are within the {\textit {Planck}} 95$ \% $ C.L. in Eq. (\ref{constraint}).
   
  \vspace{0.2cm} 
  
\begin{figure}[h!]
 \hspace{-1cm}
 \includegraphics[width=0.435\linewidth]{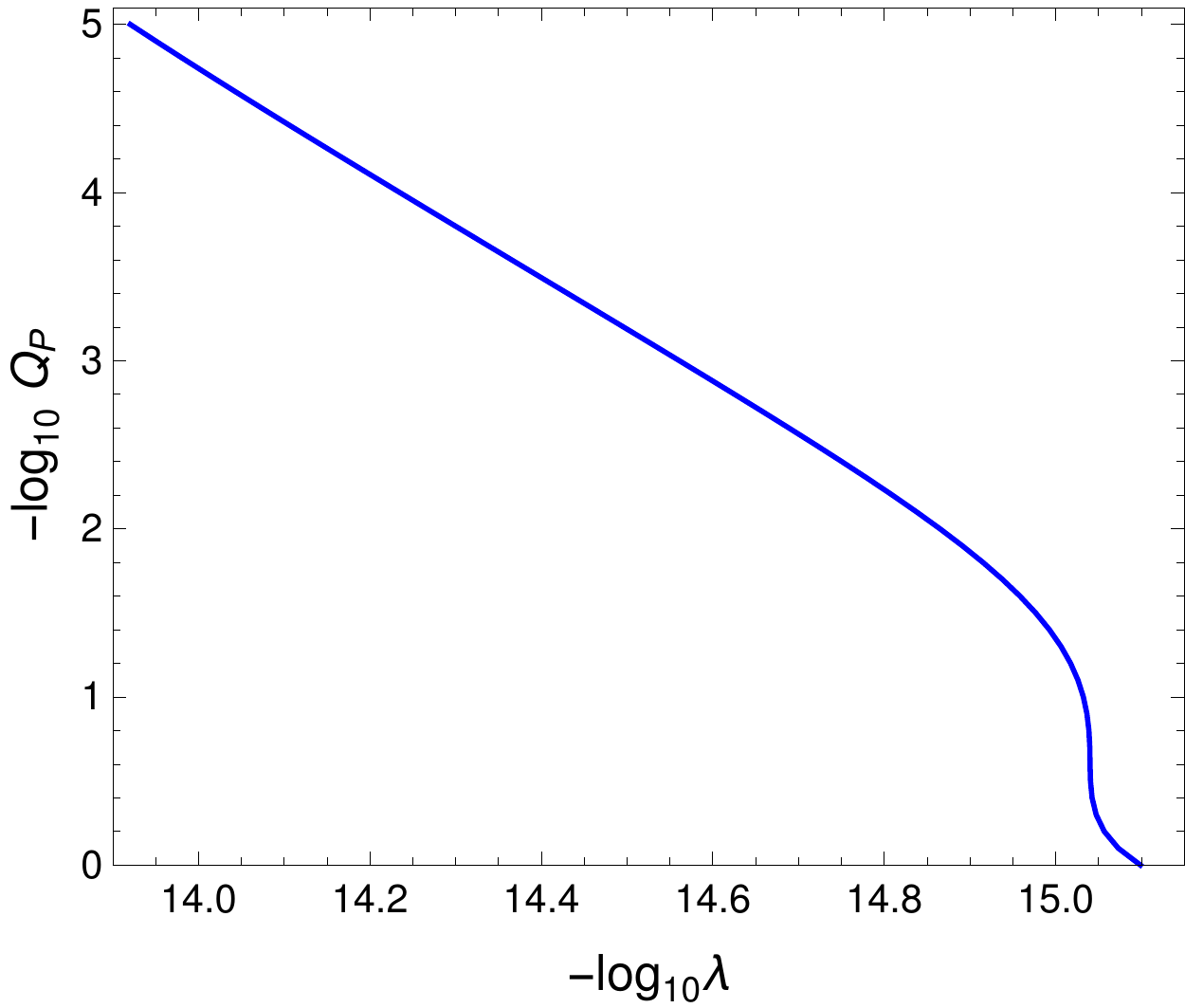}
 \includegraphics[width=0.49\linewidth]{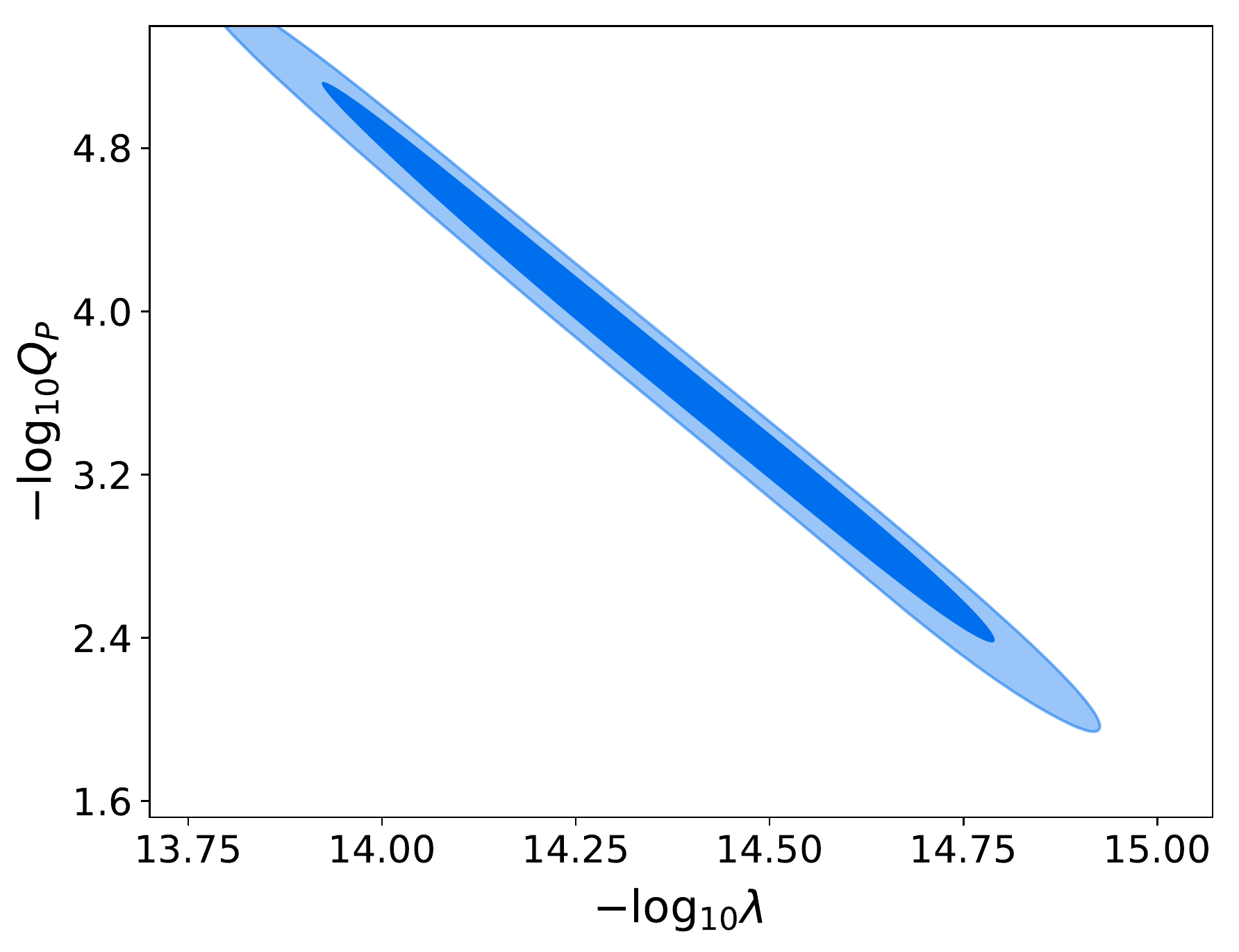}
 \includegraphics[width=0.452\linewidth]{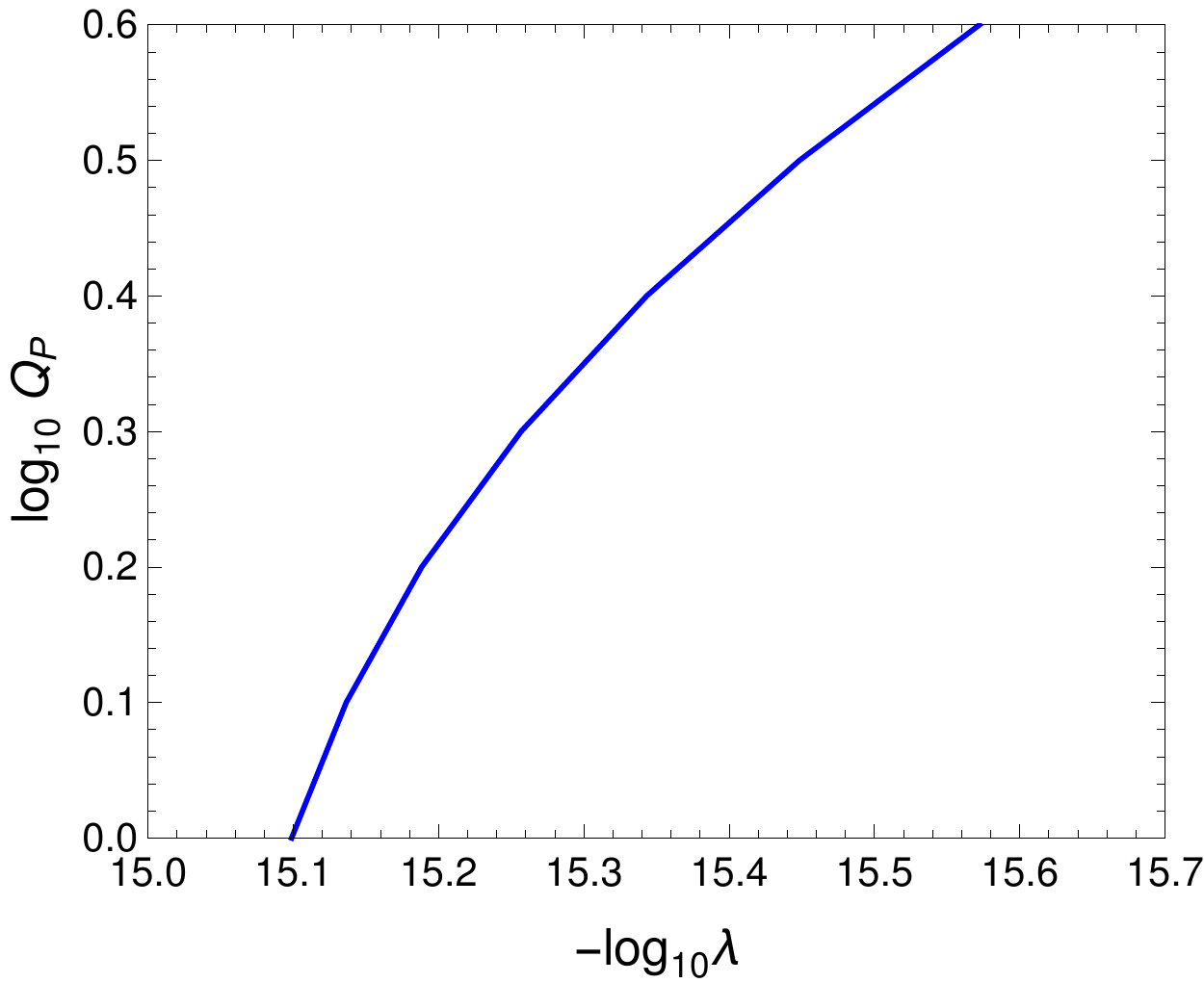}
\includegraphics[width=0.49\linewidth]{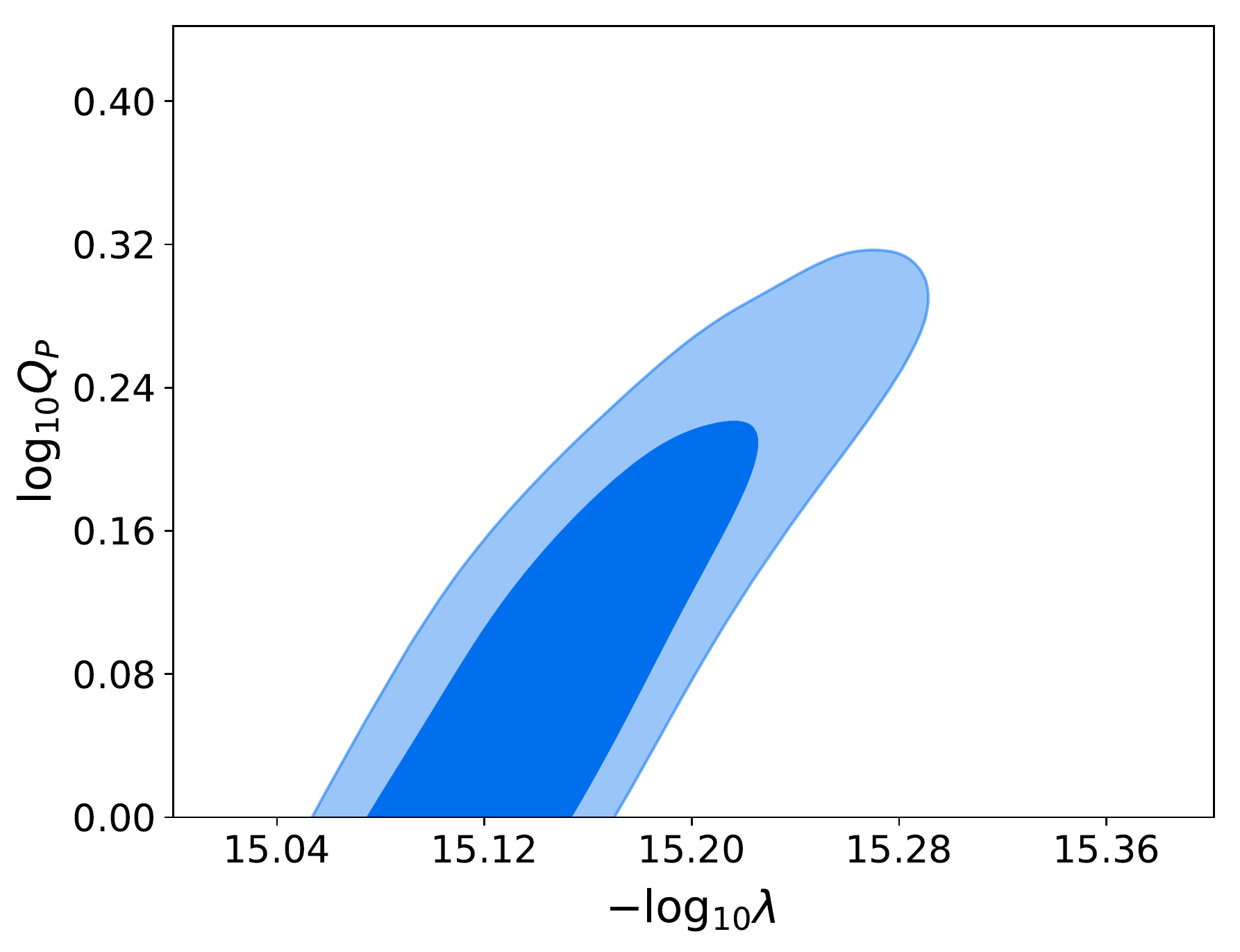}
\vspace{-0.4cm}
 \caption{Joint probability distribution of $-\log_{10} Q_P$ and $-\log_{10}\lambda$ for the case $V(\phi)=\lambda  \phi^4$, with $\Upsilon= C_T T$ in the weak (\textit{Top}) and between $\log_{10} Q_P$ and $-\log_{10}\lambda$ in the strong (\textit{Bottom}) dissipative regime. \textit{Left}:  From Mathematica with the normalisation condition $A_s=2.2\times 10^{-9}$ and \textit{Right}: from CosmoMC with $\lambda$ and $Q_P$ as parameters. $N_P=60$ for all plots.}
 \label{fig:QL} 
 \end{figure}

  In Fig. \ref{fig:QL}, we show the joint probability distribution for the $Q_P$ and $\lambda$, in the weak and strong dissipative regimes. The \textit{Left} plots are obtained in {\tt Mathematica} keeping the normalisation of the primordial power spectrum fixed at a value $A_s = 2.2\times 10^{-9}$ and the \textit{Right} plots are the contour plots with $1\sigma$ and $2\sigma$ regions obtained via {\tt CosmoMC}.
  
 \vspace{0.2cm}
 In the weak dissipative regime, $\lambda \propto Q_P^{-0.3}$ in both the {\tt Mathematica} and {\tt CosmoMC} generated plots and for the strong dissipative regime, $\lambda \propto Q_P^{-0.6}$. One sees that the behaviour of $\lambda-Q_P$ differs in the two regimes.
\\
 \subsection{Model II: $V(\phi)=\lambda\frac{\phi^6}{M_{Pl}^2}$ and $\Upsilon= C_T T$}
  The priors for the model parameters   and the mean values along with 68\% limits in the weak and the strong dissipative regimes are listed in Tables \ref{tab:wSL} and \ref{tab:sSL}.

 \begin{table}[H]\vspace{0.3cm}
 \begin{minipage}{0.5\textwidth}
\begin{tabular}{c c c} \hline\hline
 { Parameter} & { Priors}  & 68\% limits \vspace{0.1cm} \\ \hline \vspace{0.1cm}
 $\Omega_bh^2$ & [0.005,0.1]  & $   0.02157\pm 0.00013 $ 
 \vspace{0.1cm}
 	\\
 $\Omega_ch^2$ & [0.001,0.99] & $   0.12484\pm 0.00099$ 
 \vspace{0.1cm}
 	\\ 
 $100\theta$   & [0.50,10.0]  &  $  1.03989 \pm 0.00029$\vspace{0.1cm}
 	\\ 
 $\tau$ &  [0.01,0.8]   & $ 0.056 \pm 0.020$  \vspace{0.1cm}
 	\\ 
 $-\log_{10}\lambda$ & [15.4,16.6]& $ 16.07^{+0.27 }_{-0.19}$\vspace{0.1cm}
 	\\ 
 $-\log_{10}Q_P$ &  [1.8,5.4]  & $  3.54 ^{+0.68}_{-0.82}$\vspace{0.1cm}
 \\ 
 \hline\hline
\end{tabular}
\end{minipage}\hfill
\begin{minipage}{0.4\textwidth}
Mean value of $\lambda$ = $8.51 \times 10^{-17}$\\
 Mean value of $Q_P$ = $2.88 \times 10^{-4}$
 \vspace{0.3cm}
	\\
 For these values, we obtain\\ 
 $n_s=0.956$\\
 $r=0.0451$
 \end{minipage}
 	\vspace{0.1cm}
 \caption{The priors and the marginalised values along with 68\% limits for the parameters of the model  $V(\phi)=\lambda\frac{\phi^6}{M_{Pl}^2}$ with $\Upsilon= C_T T$ in the weak dissipative regime are shown here. The mean values of the model parameters and the corresponding values of $n_s$ and $r$ are also given. 
 }
\label{tab:wSL}
 \end{table}

\vspace{0.2cm}

 \begin{table}[h!]
 \begin{minipage}{0.5\textwidth}
\begin{tabular}{c c c} \hline\hline
 { Parameter} & { Priors}  & 68\% limits \vspace{0.1cm} \\ \hline \vspace{0.1cm}
 $\Omega_bh^2$ & [0.005,0.1]  & $   0.02170 \pm 0.00014 $ 
 \vspace{0.1cm}
 	\\
 $\Omega_ch^2$ & [0.001,0.99] & $   0.1206 \pm 0.0015$ 
 \vspace{0.1cm}
 \\ 
 $100\theta$   & [0.50,10.0]  &  $  1.04037 \pm 0.00030 $\vspace{0.1cm}
 \\ 
 $\tau$ &  [0.01,0.8]   & $ 0.066 \pm 0.022 $  \vspace{0.1cm}
 \\ 
 $-\log_{10}\lambda$ & [14.8,15.9]& $ 15.253 \pm 0.029$\vspace{0.1cm}
	\\ 
 $\log_{10}Q_P$ &  [0,1.5]  & $   0.596 \pm 0.048$\vspace{0.1cm}
 	\\ 
 \hline\hline
\end{tabular}
\end{minipage}\hfill
\begin{minipage}{0.4\textwidth}
Mean value of $\lambda = 5.59 \times 10^{-16}$\\
 Mean value of $Q_P$ = $3.94  $
 \vspace{0.3cm}
 \\
 For these values, we obtain\\ 
 $n_s=0.970$\\
 	$r=0.0000426$
\end{minipage}
\caption{The priors and the marginalised values along with 68\% limits for the parameters of the model  $V(\phi)=\lambda\frac{\phi^6}{M_{Pl}^2}$ with $\Upsilon= C_T T$ in the strong dissipative regime are shown here. The mean values of the model parameters and the corresponding values of $n_s$ and $r$ are also given.  
 }
\label{tab:sSL}
\end{table}
\vspace{-0.4cm}
The mean values of our $Q_P$ obtained from {\tt CosmoMC} for the weak and strong dissipative regimes lie in the allowed range of values of $Q_P$ in the $n_s$ plot in Fig. \ref{SL}.  The values of $n_s$ and $r$ obtained from the mean values of $\lambda$ and $Q_P$ are within the {\textit {Planck}} 95$ \% $ C.L. as given in Eq. (\ref{constraint}).

\begin{figure}[h!]
\centering
\includegraphics[width=0.47\linewidth]{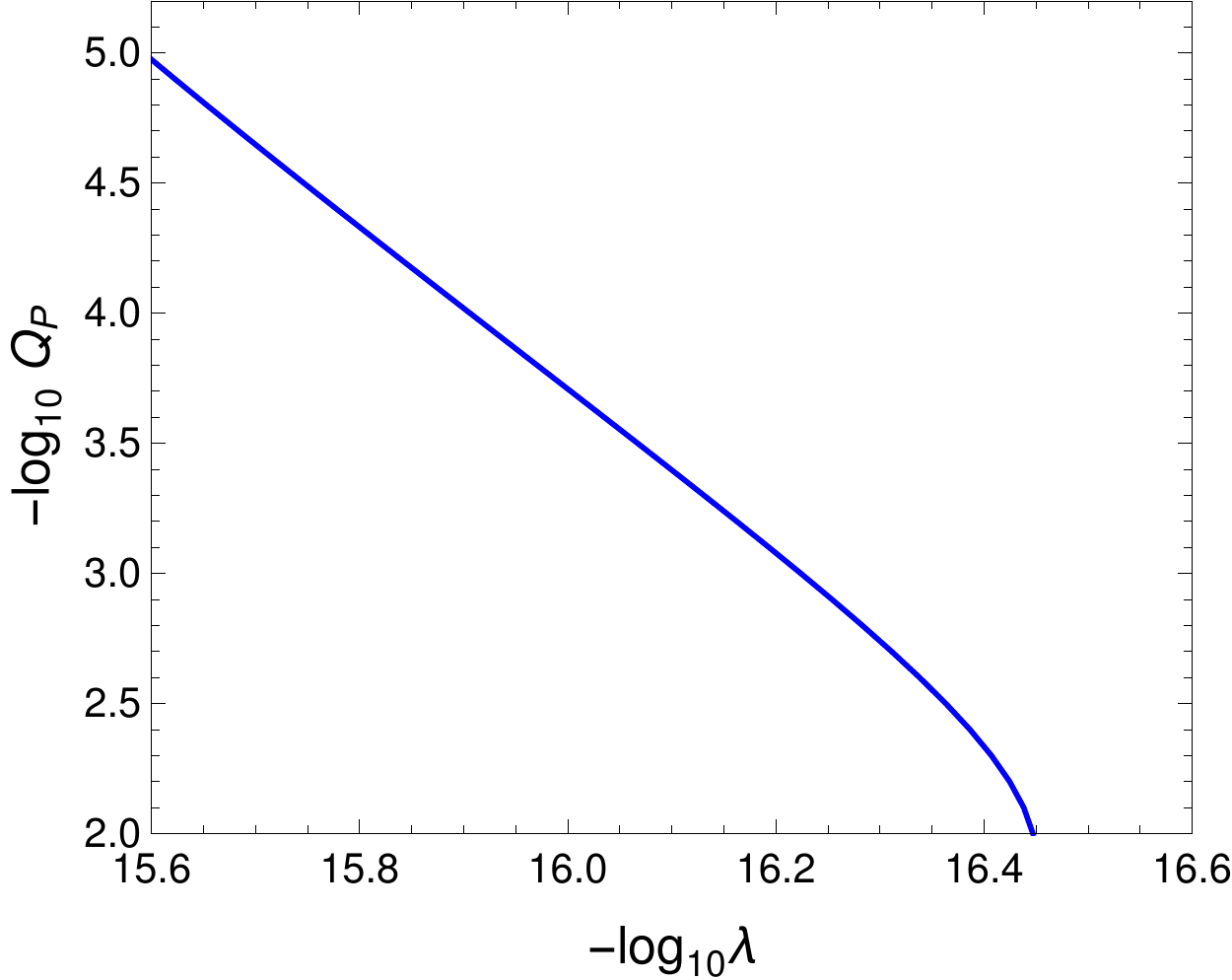}
\includegraphics[width=0.5\linewidth]{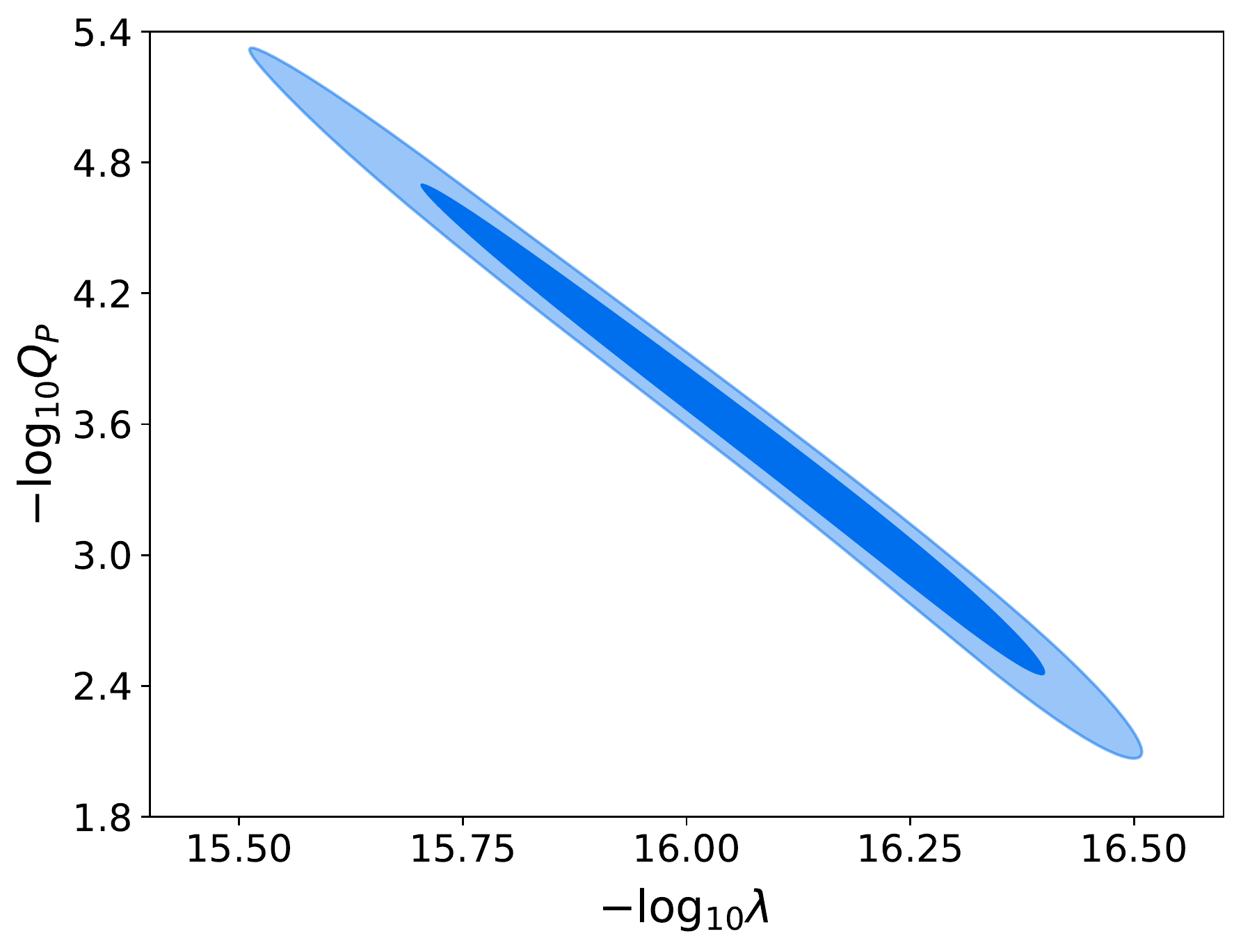}
 \includegraphics[width=0.465\linewidth]{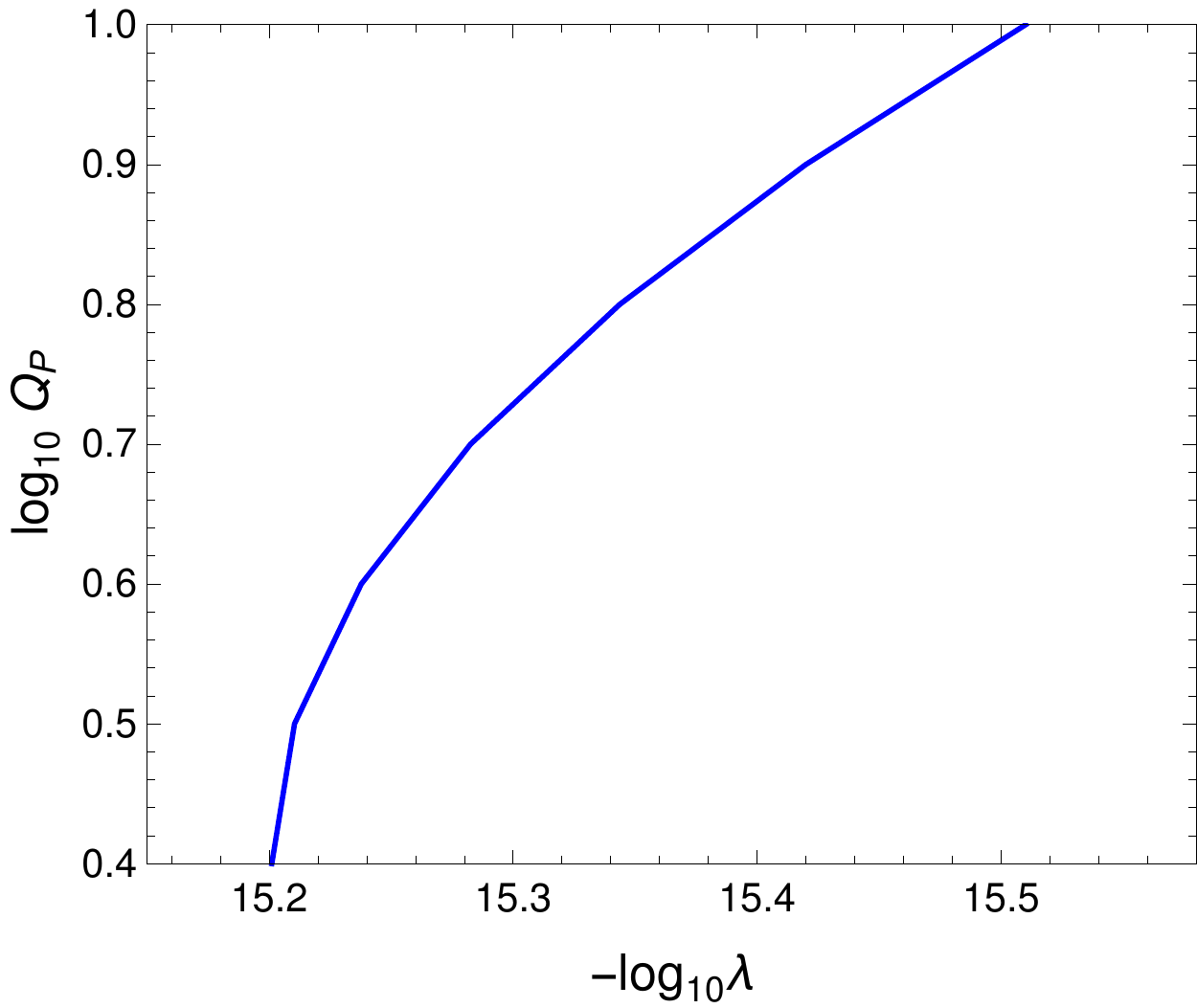}
 \includegraphics[width=0.51\linewidth]{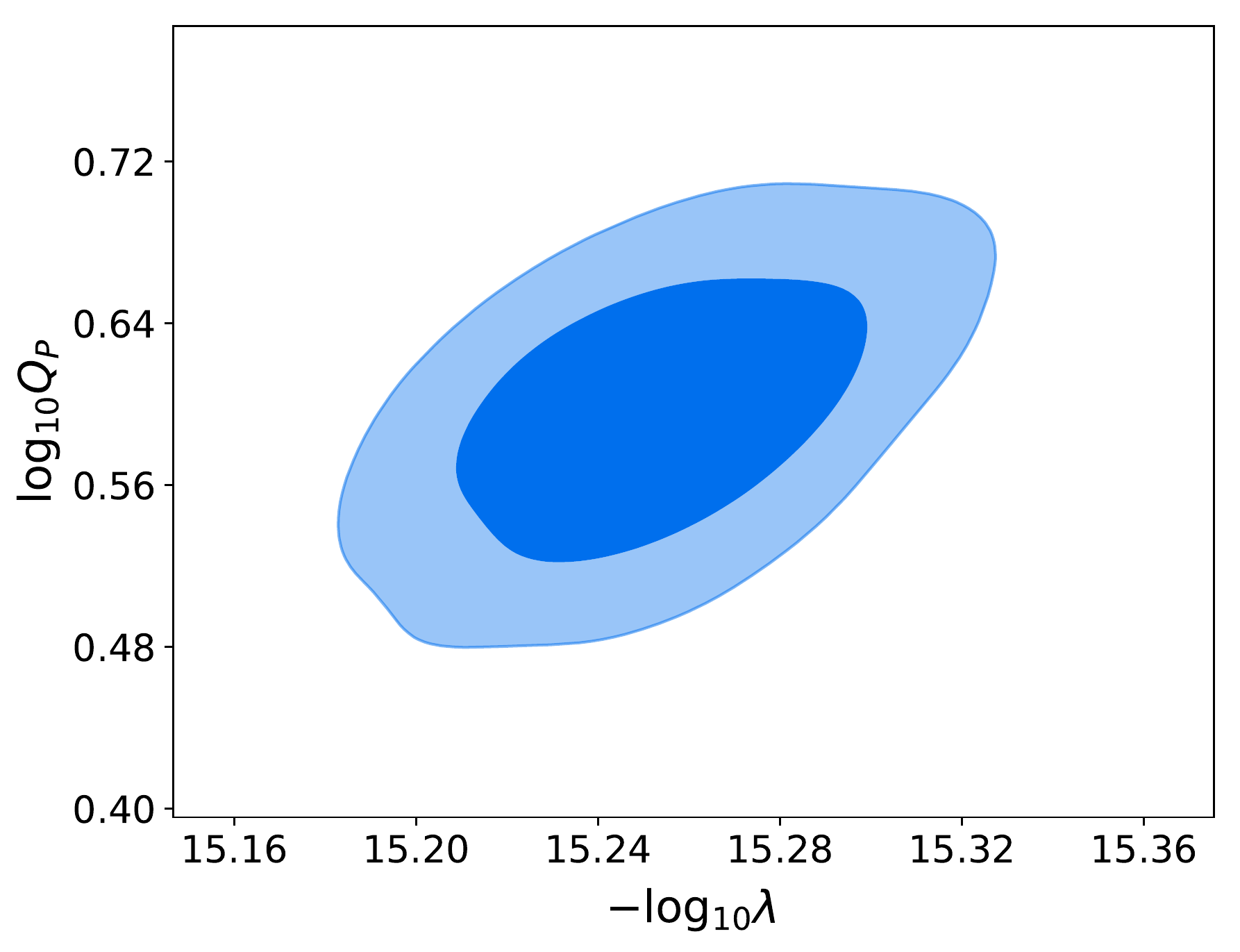}
 \vspace{0.2cm}
 \caption{Joint probability distribution of $-\log_{10}\lambda$ and $\log_{10} Q_P$ for the case $V(\phi)=\lambda\frac{\phi^6}{M_{Pl}^2}  $, with $\Upsilon= C_T T$ in the weak (\textit{Top}) and strong (\textit{Bottom}) dissipative regime. \textit{Left}:  From Mathematica with the normalisation condition $A_s=2.2\times 10^{-9}$ and \textit{Right}: from CosmoMC with $\lambda$ and $Q_P$ as  parameters. $N_P=60$ for all plots.}
 \label{fig:SL}
\end{figure}

 In Fig. \ref{fig:SL}, we show the joint probability distribution for the $Q_P$ and $\lambda$, in the weak and strong dissipative regimes.
 Here again, the \textit{Left} plots are obtained in {\tt Mathematica} and the \textit{Right} plots are the contour plots with $1\sigma$ and $2\sigma$ regions obtained via {\tt CosmoMC}.
 
 \vspace{0.2cm}
 In the weak dissipative regime, $\lambda \propto Q_P^{-0.3}$ in both the {\tt Mathematica} and {\tt CosmoMC} generated plots and for the strong dissipative regime, $\lambda \propto Q_P^{-0.4}$. Here also, we can see that the behaviour of $\lambda-Q_P$ differs in the two regimes.
 \\
  \subsection{Model III: $V(\phi)=\lambda\frac{\phi^6}{M_{Pl}^2}$ and $\Upsilon= C_{\phi} \frac{T^3}{\phi^2}$}
 
\begin{figure}[h!]
   \vspace{-0.3cm}
   	\includegraphics[width=0.95\linewidth]{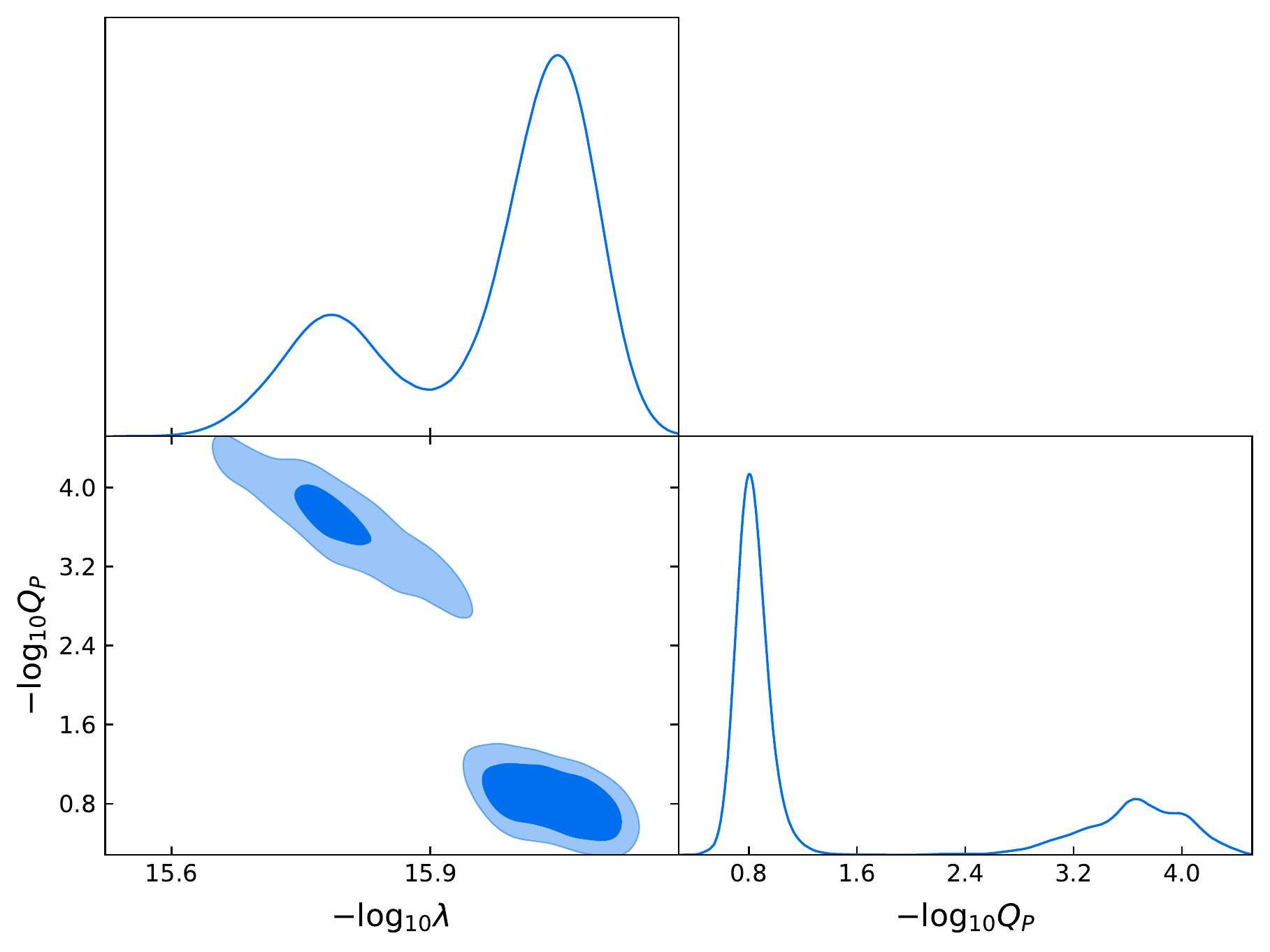}
   	 \vspace{-0.2cm}
   	\caption{Joint probability distribution of $-\log_{10}\lambda$ and $-\log_{10} Q_P$ for the case $V(\phi)=\lambda\frac{\phi^6}{M_{Pl}^2}  $, with $\Upsilon= C_{\phi} \frac{T^3}{\phi^2}$ in the weak dissipative regime. The figure shows that there are two convergence regions, however the probability for $-\log_{10} Q_P$ near 0 is higher as can be seen from the height of the peak.}
   	\label{SC2}
   \end{figure}

 As previously mentioned, in this case, we consider only the weak dissipative regime as the $n_s$ values in the strong dissipation regime overshoot the \textit{Planck } 2015 allowed region. We obtain two convergence regions in our analysis with {\tt CosmoMC}. This is shown in Fig. \ref{SC2}. However the region with $-\log Q_P$ close to 0 has a higher probability, as shown by the height of the probability distribution peak. Therefore we redo our analysis by restricting the priors such that we only obtain the mean value of $Q_P$ in the more probable region. We write the priors for the parameters and the mean values along with 68\% limits in Table \ref{tab:wSC}.  \\
  
 \begin{table}[h!]
 \begin{minipage}{0.5\textwidth}
  \begin{tabular}{c c c} \hline\hline
  { Parameter} & { Priors}  & 68\% limits \vspace{0.1cm} \\ \hline \vspace{0.1cm}
  $\Omega_bh^2$ & [0.005,0.1]  & $   0.02170\pm 0.00013 $ 
  \vspace{0.1cm}
  \\
  $\Omega_ch^2$ & [0.001,0.99] & $   0.1207\pm0.0014$ 
  \vspace{0.1cm}
  \\ 
  $100\theta$   & [0.50,10.0]  &  $  1.04036\pm0.00030 $\vspace{0.1cm}
  	\\ 
  $\tau$ &  [0.1,0.8]   & $ 0.061\pm0.023 $  \vspace{0.1cm}
  	\\ 
  $-\log_{10}\lambda$ & [15.8,17.0]& $16.064 \pm 0.38$\vspace{0.1cm}
  \\ 
  $-\log_{10}Q_P$ &  [0,1.5]  & $  0.799 ^{+0.068}_{-0.10}$\vspace{0.1cm}
  \\ 
  \hline\hline
  \end{tabular}
  		
\end{minipage}\hfill
\begin{minipage}{0.4\textwidth}
 Mean value of $\lambda$ =$8.63\times 10^{-17}$ \\
  Mean value of $Q_P$ =$0.1588$ 
  \vspace{0.3cm}
  \\
  For these values, we obtain\\ 
  $n_s=0.969$\\
  $r=0.00480$
 \end{minipage}
 \vspace{0.1cm}
 \caption{The priors and the marginalised values along with 68\% limits for the parameters of the model  $V(\phi)=\lambda  \frac{\phi^6}{M_{Pl}^2}$ with 
 $\Upsilon=C_\phi \frac{T^3}{\phi^2} $ in the weak dissipative regime are shown here. The mean values of the model parameters and the corresponding values of $n_s$ and $r$ are also given.   
 }
 \label{tab:wSC}
\end{table}

 \begin{figure}[h!]
 \vspace{0.5cm}
\includegraphics[width=0.475\linewidth]{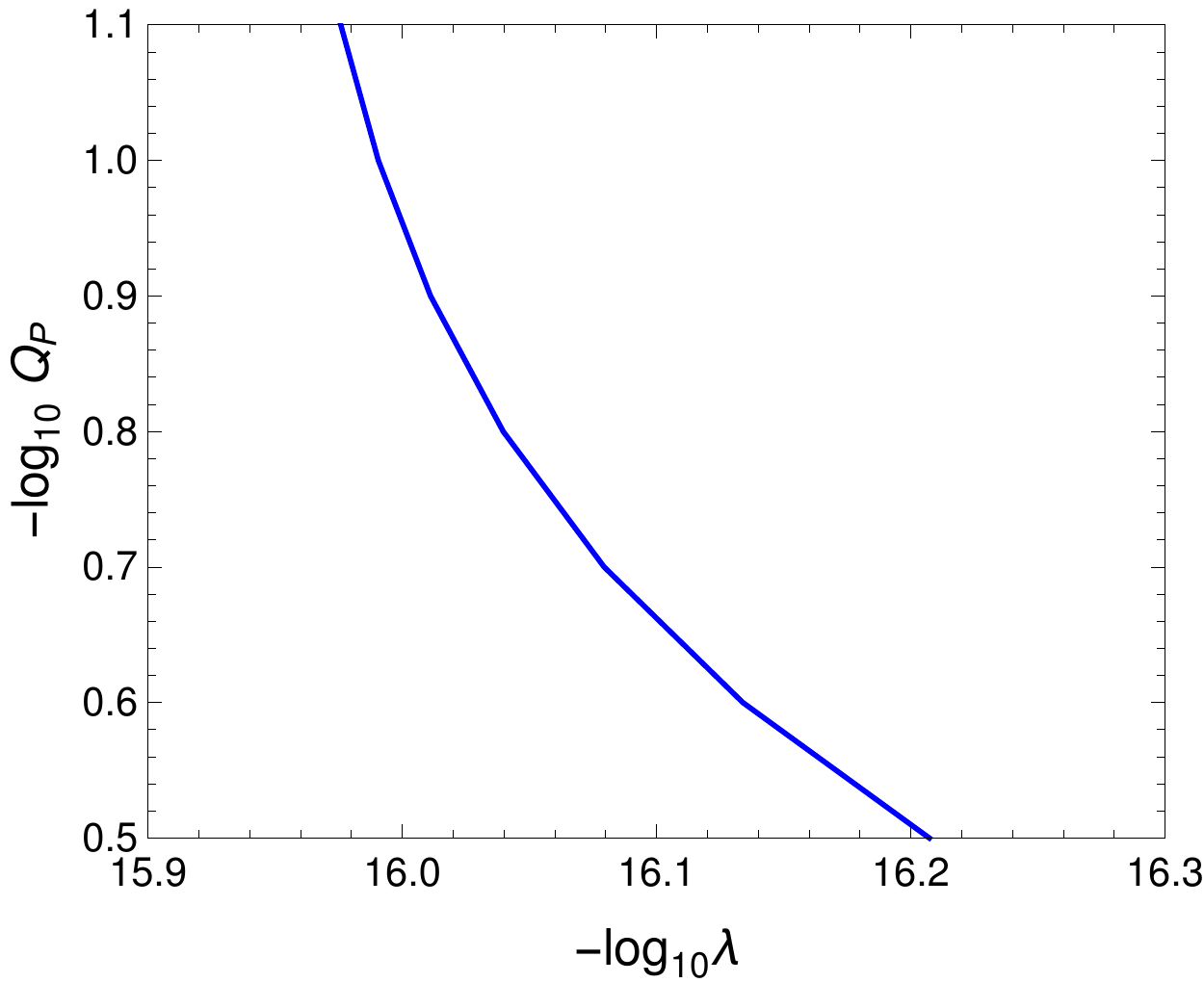}
\includegraphics[width=0.5\linewidth]{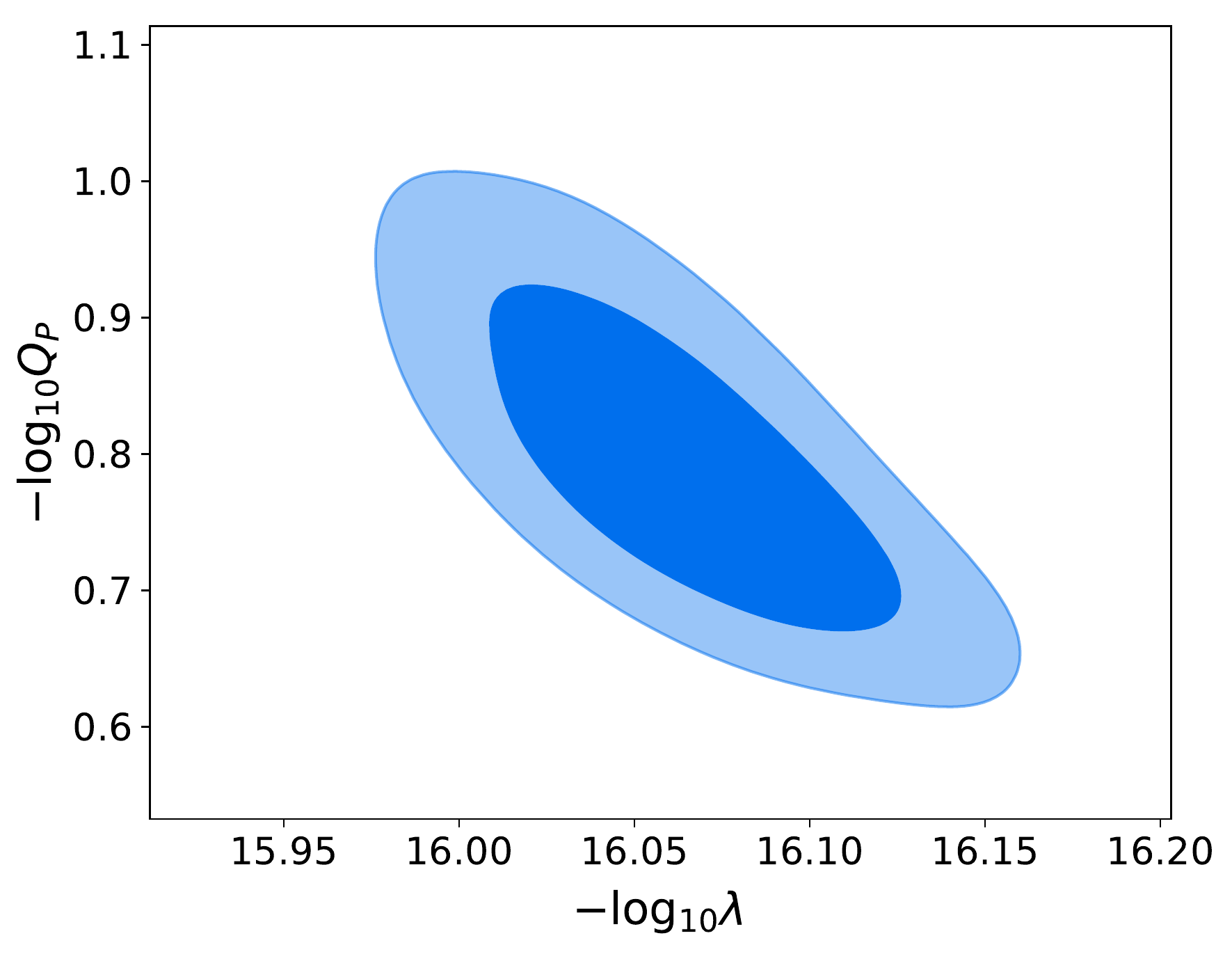}
\caption{Joint probability distribution of $-\log_{10}\lambda$ and $-\log_{10} Q_P$ for the case $V(\phi)=\lambda\frac{\phi^6}{M_{Pl}^2}  $, with $\Upsilon= C_{\phi} \frac{T^3}{\phi^2}$ in the weak dissipative regime. \textit{Left}:  From Mathematica with the normalisation condition ($A_s=2.2\times 10^{-9}$) and \textit{Right}: from CosmoMC with $\lambda$ and $Q_P$ as  parameters. $N_P=60$ for all plots.}
\label{fig:SC}
 \end{figure}
 
 The mean values of our $Q_P$ obtained from {\tt CosmoMC} for the weak dissipative regime lie in the allowed range of values of $Q_P$ in the $n_s$ plot in Fig. \ref{SC}.  The value of $n_s$ obtained from the mean values of $\lambda$ and $Q_P$ is within the {\textit {Planck}} 68$ \% $ C.L. in Eq. (\ref{constraint}) while the value of $r$ is consistent with the {\it Planck} bounds.

 \vspace{0.2cm}
 
 In Fig. \ref{fig:SC}, we show the joint probability distribution for the $Q_P$ and $\lambda$, in the weak  dissipative regime. Here again, the \textit{Left} plot is obtained in {\tt Mathematica} and the \textit{Right} plot is the contour plot with $1\sigma$ and $2\sigma$ regions obtained via {\tt CosmoMC}.
 
 \vspace{0.2cm}
 The relation between $\lambda-Q_P$ goes as $\lambda \propto Q_P^{-0.4}$ in the weak dissipative regime, which is similar in both the {\tt Mathematica} and {\tt CosmoMC} analyses.

 \section{Conclusion}
 \label{conc}
  In standard cold inflation, the inflaton's couplings with other fields are ineffective during the inflationary phase and as a result the Universe is supercooled during inflation. Warm inflation provides an alternate picture of inflation where radiation production takes place concurrent to the inflationary phase and the Universe has a temperature during inflation. With the current measurements of temperature anisotropies in the Cosmic Microwave Background radiation, various monomial potentials of standard cold inflation are ruled out from the $n_s-r$ parameter space. However, certain monomial potentials are still viable models in the context of warm inflation.
 
 \vspace{0.2cm}
 As an extension to our previous study on estimating the parameters of warm inflationary $\lambda\phi^4$ potential with a cubic dissipation coefficient, here we have discussed the $\lambda \phi^4$ potential model of warm inflation with a linear dissipation coefficient
 and the $\lambda\phi^6$ potential model of warm inflation with linear and cubic dissipation coefficients. We have shown the dynamics of the inflaton and radiation during warm inflation. We parameterize the primordial power spectrum for our models and study its dependence on the model parameters, namely, $\lambda$ signifying the inflaton self coupling, and $Q_P$, representing the dissipation parameter arising because of the inflaton's couplings to the other fields. 
 After a preliminary analysis using Mathematica, we
 perform a Markov Chain Monte Carlo (MCMC) analysis over these model parameters  and four standard parameters using the {\tt CosmoMC} numerical code. This provides us the mean values  of all the parameters with $68\%$ confidence limits for all the models we consider. For the mean values of parameters, we also estimate the values for the spectral tilt, $n_s$ and the tensor-to-scalar ratio $r$
 and find them to be consistent with the {\textit Planck}  allowed ranges for all the models we consider. We also obtain the joint probability contours with $1 \sigma$ and $2 \sigma$ regions for $\lambda$ and $Q_P$ with the {\tt GetDist} package.   For all the models, we approximate a linear fit to the joint probability curve of $\log\lambda$ and $\log Q_P$ and find a relation between them. An estimation of the parameters of warm inflation models and the relations between them is important from the perspective of model building.
  
   \vspace{0.2cm}
  For the weak dissipative region, our values of $r$ for the mean values of the model parameters is within the sensitivity of order $10^{-3}$ of the next generation of ground-based and satellite-based  CMB polarisation experiments \cite{Abazajian:2016yjj,Ishino:2016izb,Creminelli:2015oda}. However, for the strong dissipative region, the corresponding values of $r$ are lower than the sensitivity of these future experiments.
  It has been argued that  lensing of intensity fluctuations in
  the 21-cm signal from atomic hydrogen in the dark ages can in principle provide a probe of inflationary gravitational waves down to a sensitivity of $10^{-9}$ for $r$ \cite{Book:2011dz}.  However such measurements would be challenging and require a futuristic experiment.  
  Experiments have also been proposed to measure the redshifted 21-cm hydrogen line and use interferometry to obtain a sensitivity of $10^{-3}$ on the Earth, and a sensitivity of $10^{-6}$ with an  array of detectors covering a large area of the Moon's surface \cite{Ansari:2018ury}.  Thus while the warm inflation models studied here for the weak dissipative regime can be further investigated via CMB polarisation experiments in the near future, one will have to wait much longer to test these models in the strong dissipative regime.\\  
\vspace{0.2cm}
\\
\textbf{Acknowledgements}
\vspace{0.2cm}
\\
The results presented in the paper are based on the computations using Vikram-100, the 100TFLOP HPC Cluster at Physical Research Laboratory, Ahmedabad, India.
We would like to thank Prakrut Chaubal and Jayanti Prasad for their help in CosmoMC related issues. We would also like to thank Gaurav Goswami, Namit Mahajan and Arvind Kumar Mishra for  valuable discussions and suggestions.

\appendixpage
 \appendix

   \section{Dissipation parameter at the end of inflation, 
   	and the integral function
   	}
   \label{sol}
   \subsection*{For $V(\phi)=\lambda  \phi^4$ with  $\Upsilon= C_T T$}
   \vspace{0.2cm}
   The positive real solution to Eq. (\ref{qe1}) is given by
  \be
  Q_e(\lambda,C_T)=\frac{\frac{2}{3}^{1/3}Y}{(9Y
   + \sqrt{3}\sqrt{27Y^2-4Y^3})^{1/3}}+\frac{(9Y
   + \sqrt{3}\sqrt{27Y^2-4Y^3})^{1/3}}{{2}^{1/3}{3}^{2/3}}
   \ee
    where $Y= \frac{1}{12^3}\frac{4C_\phi^4}{9A\lambda }.$
    In this way, $Q_e$ is expressed in terms of $\lambda$ and $C_T$.
   \vspace{0.2cm}
   \\
   The integral function in Eq. (\ref{Np}) is given as
    \be
    F(Q)=\frac{1}{Z}\frac{-3 (1 + Q)^{\frac{1}{3}} (2 + 3\,\, {_2F_1}(1, 1; 5/3;-1/Q))}{2 Q}
    \ee
    where $Z= \frac{1}{24}\left(\frac{4C_T^4}{9A\lambda }\right)^{1/3}$ and ${_2F_1}(a,b;c;z)$ is the hypergeometric function.
  \\
  \vspace{0.3cm}
  \subsection*{For  $V(\phi)=\lambda\frac{\phi^6}{M_{Pl}^2}$ with  $\Upsilon= C_T T$}
  \vspace{0.2cm}
  The positive real solution to Eq. (\ref{qe2}) is given by
   \begin{align}
  Q_e(\lambda,C_T)&=\frac{Y}{3}+\frac{2^{1/3}(6Y+Y^2)}{3(27 Y + 18 Y^2 + 2 Y^3+3\sqrt{3}\sqrt{(27Y^2+4Y^3})^{1/3}}
  \nonumber\\  
  &+\frac{(27 Y + 18 Y^2 + 2 Y^3+3\sqrt{3}\sqrt{(27Y^2+4Y^3})^{1/3}}{ 2^{1/3}\,3}
  \end{align} where $Y= \frac{1}{30^4}\frac{ C_T^4 8\pi}{A \lambda }.$
  In this way, $Q_e$ is expressed in terms of $\lambda$ and $C_T$.\\
  \vspace{0.3cm}
  \\
  The integral function in Eq. (\ref{Np}) is given as
  \be
  F(Q)=\frac{1}{Z}\frac{4 ((1 + Q)^{\frac{1}{2}} -4 \,\, Q \,\,{_2F_1}(1/4, 1/2; 5/4;-Q))}{Q^{\frac{3}{4}}}
  \ee
  where $Z= \frac{\pi}{6}\left(\frac{C_T^4}{A\lambda (8\pi)^3 }\right)^{1/4}.$
  \\
   \subsection*{For $V(\phi)=\lambda  \frac{\phi^6}{M_{Pl}^2}$ with  $\Upsilon=C_\phi \frac{T^3}{\phi^2} $}
   \vspace{0.2cm}
   The positive real solution to Eq. (\ref{qe3}) is given by
   \be
   Q_e(\lambda,C_\phi)=\frac{1}{3}\left(-2 + \frac{2^{1/3}}{(2 + 27 Y + 3 \sqrt{3} \sqrt{4 Y + 27 Y^2})^{
   1/3}} + \frac{(2 + 27 Y + 3 \sqrt{3} \sqrt{4 Y + 27 Y^2})^{
   1/3}}{2^{1/3}}\right)
   \ee
   where $Y= \frac{1}{10^4}\frac{\lambda C_\phi^4 }{8\pi A^3}.$
   In this way, $Q_e$ is expressed in terms of $\lambda$ and $C_\phi$.
   \\
   \vspace{0.3cm}
    \\
    The integral function in Eq. (\ref{Np}) is given as
    \be
    F(Q)=\frac{1}{Z}\frac{-4+8  Q-8/3 Q(1+Q) \,\, {_2F_1}(1,5/4; 7/4; -Q)}{ Q^{1/4} (1+Q)^{1/2}}
    \ee
    where $Z= \frac{1}{16}\left(\frac{\lambda C_\phi^4}{8\pi A^3 }\right)^{1/4}.$
    \\

   \section{Expression of the spectral index, $n_s$} 
     \label{ns}
    
    \subsection*{For $V(\phi)=\lambda  \phi^4$ with  $\Upsilon= C_T T$}
    \vspace{0.2cm}
    \begin{align}
    n_s &=1 - \frac{9\epsilon_H(1+Q_P)}{(1- \epsilon_H)(3 + 5\,Q_P)} + X\left( \frac{2H_P}{T_P}\exp\left(\frac{H_P}{T_P}\right)n_P^2+\frac{T_P}{H_P}\frac{2 \sqrt{3} \pi Q_P}{\sqrt{3 + 4 \pi Q_P}}\, \frac{6+6\pi Q_P}{3+4\pi Q_P}\right)  
    \nonumber\\  
    &+\frac{3\epsilon_H}{(1-\epsilon_H)}\frac{Q_P(1+Q_P)}{3+5\,Q_P}\,\frac{0.042827 \,Q_P^{1.315} + 0.45694 \,Q_P^{0.364}}{1 + 0.0185 \,Q_P^{2.315} + 0.335 \,Q_P^{1.364}} 
    \label{nsQL}
    \end{align}
    \vspace{0.2cm}
    \\
    where
    $X=\frac{1}{\left[1+2n_P+\left(\frac{T_P}{H_P}\right)
    \frac{2\sqrt{3}\pi Q_P}{\sqrt{3+4\pi Q_P}}\right]}\,\, \frac{3\epsilon_H(1+Q_P)}{(1- \epsilon_H)(3 + 5\,Q_P)} $ and $\epsilon_H$ is evaluated at $k=k_P.$ 
    \vspace{0.5cm}
    \\
     \subsection*{For  $V(\phi)=\lambda\frac{\phi^6}{M_{Pl}^2}$ with  $\Upsilon= C_T T$}
     \vspace{0.2cm}
     \begin{align}
   n_s &=1 - \frac{8\epsilon_H(1+Q_P)}{(1- \epsilon_H)(3 + 5\,Q_P)} + X\left( \frac{2H_P}{T_P}\exp\left(\frac{H_P}{T_P}\right)n_P^2+\frac{T_P}{H_P}\frac{2 \sqrt{3} \pi Q_P}{\sqrt{3 + 4 \pi Q_P}}\, \frac{6+6\pi Q_P}{3+4\pi Q_P}\right)  \nonumber\\  
   &+\frac{8\epsilon_H}{3(1-\epsilon_H)}\frac{Q_P(1+Q_P)}{3+5\,Q_P}\,\frac{0.042827 \,Q_P^{1.315} + 0.45694 \,Q_P^{0.364}}{1 + 0.0185 \,Q_P^{2.315} + 0.335 \,Q_P^{1.364}}  
     \label{nsSL}
     \end{align}
     \vspace{0.2cm}
     \\
    where $X=\frac{1}{\left[1+2n_P+\left(\frac{T_P}{H_P}\right)
    \frac{2\sqrt{3}\pi Q_P}{\sqrt{3+4\pi Q_P}}\right]}\,\, \frac{8\epsilon_H(1+Q_P)}{3(1- \epsilon_H)(3 + 5\,Q_P)} $ and $\epsilon_H$ is evaluated at $k=k_P.$ 
    \vspace{0.5cm}
    \\
    \subsection*{For $V(\phi)=\lambda  \frac{\phi^6}{M_{Pl}^2}$ with  $\Upsilon=C_\phi \frac{T^3}{\phi^2} $}
    \vspace{0.2cm}
    \begin{align}
      n_s &=1 - \frac{8\epsilon_H(1+5\,Q_P)}{3(1- \epsilon_H)(1 + 7\,Q_P)} + X\left( \frac{2H_P}{T_P}\exp\left(\frac{H_P}{T_P}\right)n_P^2+\frac{T_P}{H_P}\frac{2 \sqrt{3} \pi Q_P}{\sqrt{3 + 4 \pi Q_P}} \left( 1+\frac{2(1+Q_P)(3+2\pi Q_P)}{(1+3\,Q_P)(3+4\pi Q_P)}\right) \right)  \nonumber\\  
      &+\frac{8\epsilon_H}{3(1-\epsilon_H)}\frac{Q_P(1+Q_P)}{1+7\,Q_P}\,\,\frac{9.693026 \,Q_P^{0.946} + 0.54991 \,Q_P^{3.330}}{1 + 4.981\,Q_P^{1.946} + 0.127\,Q_P^{4.330}} 
      \label{nsSC}
      \end{align}
     where
     $X=\frac{1}{\left[1+2n_P+\left(\frac{T_P}{H_P}\right)
     	\frac{2\sqrt{3}\pi Q_P}{\sqrt{3+4\pi Q_P}}\right]}\,\, \frac{4\epsilon_H(1+3Q_P)}{3(1- \epsilon_H)(1+7Q_P)} $ and $\epsilon_H$ is evaluated at $k=k_P.$ 
      \\
      
 \end{document}